# Grayscale-to-color: Single-step fabrication of bespoke multispectral filter arrays

**Calum Williams**[1,2†]**, George S.D. Gordon**[1]**, Sophia Gruber**[1,2]**, Timothy D. Wilkinson**[1]**, Sarah E. Bohndiek**[2,3*]

[1] *Electrical Engineering Division, Department of Engineering, University of Cambridge, JJ Thomson Avenue, Cambridge, CB3 0FA, UK*

[2] *Department of Physics, Cavendish Laboratory, University of Cambridge, JJ Thomson Avenue, Cambridge, CB3 0HE, UK*

[3] *Cancer Research UK Cambridge Institute, University of Cambridge, Robinson Way, Cambridge, CB2 0RE, UK*

\* *corresponding author:* [seb53@cam.ac.uk](mailto:seb53@cam.ac.uk)
† *corresponding author:* [cw507@cam.ac.uk](mailto:cw507@cam.ac.uk)

# Abstract

Conventional cameras, such as in smartphones, capture wideband red, green and blue (RGB) spectral components, replicating human vision. Multispectral imaging (MSI) captures spatial and spectral information beyond our vision but typically requires bulky optical components and is expensive. Snapshot multispectral image sensors have been proposed as a key enabler for a plethora of MSI applications, from diagnostic medical imaging to remote sensing. To achieve low-cost and compact designs, spatially variant multispectral filter arrays (MSFAs) based on thin-film optical components are deposited atop image sensors. Conventional MSFAs achieve spectral filtering through either multi-layer stacks or pigment, requiring: complex mixtures of materials; additional lithographic steps for each additional wavelength; and large thicknesses to achieve high transmission efficiency. By contrast, we show here for the first time a single-step grayscale lithographic process that enables fabrication of bespoke MSFAs based on the Fabry-Perot resonances of spatially variant metal-insulator-metal (MIM) cavities, where the exposure dose controls insulator (cavity) thickness. We demonstrate customizable MSFAs scalable up to N-wavelength bands spanning the visible and near-infrared with high transmission efficiency (~75%) and narrow linewidths (~50 nm). Using this technique, we achieve multispectral imaging of several spectrally distinct target using our bespoke MIM-MSFAs fitted to a monochrome CMOS image sensor. Our unique framework provides an attractive alternative to conventional MSFA manufacture, by reducing both fabrication complexity and cost of these intricate optical devices, while increasing customizability.

Complementary metal-oxide-semiconductor (CMOS) image sensors are low cost and compact, implemented in a plethora of applications from digital photography to medical imaging.[1–3] Such image sensors are typically composed of millions of individually addressable pixels containing silicon photodiodes. To resolve wavelength-specific information, spatially variant and spectrally distinct optical filters are implemented. These color filter arrays (CFAs) are deposited in mosaic-like patterns atop the image sensor with a pitch matched to the pixel size. The most widespread CFA is the Bayer filter[4], which consists of red, green and blue (RGB) color filters.

In recent years, more complex mosaics incorporating additional spectral filters (wavelength bands) referred to as multispectral filter arrays (MSFAs) have been proposed to enable multi-and hyper-spectral imaging systems.[2,3,5,6] Conventional CFAs/MSFAs, however, are composed of either absorptive dyes/pigments, one for each wavelength, or multi-layer one-dimensional Bragg stacks, comprising a different combination of alternating dielectric materials for each wavelength.[3,7–9] Both approaches are cumbersome from a fabrication point of view: for $N$-wavelength bands the pigment approach requires $N$-lithographic steps and $N$-different dyes; while the Bragg approach (structural color) requires $N$-lithographic/hard-mask steps and $N$ different stack architectures.[3,7,9] The requirement for carefully aligned lithographic steps in a non-standard process, combined with the continual shrinking of pixel dimensions for higher resolutions and more complex mosaic patterns for added wavelength bands is extremely challenging for these existing MSFA fabrication methods, limiting widespread adoption.

To overcome these limitations, alternate methodologies for spectral filtering such as plasmonic[8,10–18] and high-index dielectric nanostructure arrays/metasurfaces have been proposed and investigated.[19,20] These allow tuning of electric and magnetic resonances respectively to create wavelength and polarization selectivity. Other design schemes include nanostructured diffractive arrays[21] and ultrathin dielectric coatings.[22] Such approaches overcome the need for multiple lithographic steps as they are monolithic. Unfortunately, all of these alternative approaches generally suffer from low transmission efficiencies, broad full-width-half-maximums (FWHMs) (i.e. poor wavelength selectivity) and require ultra-high resolution (sub-pixel) lithographic patterning to achieve the respective spectral filtering result, thus preventing widespread adoption as the constituent of MSFAs. Metal-insulator-metal (MIM) optical cavities have been widely shown to provide narrowband filters with high transmission efficiencies, offering a promising alternative for MSFA fabrication.[23–25] MIM-based optical filter material compositions can be deposited in one processing step, however typically each thickness is fixed.[9,25] As a result, conventional MIM-MSFAs are fabricated through iterative 'step-and-repeat' processes, limiting the number of wavelength bands.

Here, we present for the first time a novel, versatile and scalable approach based on grayscale lithography for producing highly efficient, narrowband and easily customizable MIM-MSFAs using a single lithographic step. We use grayscale electron beam lithography (G-EBL) to generate spatially variant 3D cavities through dose-modulated exposure schemes.[26] The molecular weight of the resist is modified through exposure dose, thus making the rate of development a function of dose. For a grayscale dose profile, the remaining resist thickness (post-development) depends on the dose and development time. By utilizing the 3D profile resist as the insulator material (cavity) in a MIM optical filter system, spatially dependent 3D MIM structures that act as highly efficient MSFAs with high transmission efficiency and narrow FWHM are fabricated. We demonstrate a diverse range of MSFA devices with varying mosaic complexities from the ubiquitous Bayer filter to higher order (3x3, 4x3, etc.) MSFAs used in conjunction with a monochrome image sensor for multispectral imaging. The optical performance, customizability and fabrication over relatively large areas surpasses the current state-of-the-art[8,10,19,20,11–18], representing a key enabling step towards widespread adoption of multispectral image sensors.

# Results

Our approach to the generation of visible color (spectral filtering) from grayscale dose modulation is depicted schematically in **Fig. 1**. Custom MSFAs (**Fig. 1a**) were created using dose-modulated exposure schemes to control local solubility of the resist (**Fig. 1b**). To achieve this, the 'resist sensitivity' was characterized so that a grayscale dose pattern could be applied to produce a physical three-dimensional resist profile.[26] During resist development, different filter thicknesses, and hence wavelength selectivity, could be determined on a per-pixel basis by the total energy delivered to the resist volume.

We first performed electromagnetic simulations[27] of the transmission response of a continuous MIM cavity with non-dispersive insulator (resist; n ≈ 1.65) separating the Ag (26 nm) mirrors, with addition of a 12 nm $MgF_2$ encapsulation layer (**Fig. 2a**, Supplementary Note 1). These thicknesses were determined as a tradeoff between transmission efficiency and FWHM (Supplementary Figs 1, 2). As the insulator thickness ($z$) increases, the optical path length increases and the spectral position of the mode red-shifts accordingly. Moreover, multiple transmission peaks are excited for thicker insulator layers corresponding to the additional higher order Fabry-Perot (FP)-type[23,25] modes of the system.

MIM-based MSFAs with these simulated characteristics were then fabricated experimentally. Negative-tone MaN-2400 series e-beam resist was utilized for this study due to its high-resolution capability for EBL in combination with relatively high sensitivity. An array of 5 µm (x–y dimensions) square pixels were assigned increasing dose values so that post G-EBL (with development time kept constant) each pixel had a different final thickness (z). The layers of the final MIM structure consisted of e-beam resist with two 26 nm Ag mirrors and a 12 nm $MgF_2$ encapsulation layer (see *Methods*). Experimental optical transmission spectra were recorded for each of the dose modulated (15–55 µC $cm^{-2}$) pixels (**Fig. 2b** i), with final thickness values (**Fig. 2b** ii) confirmed using an atomic force microscope (AFM). The resultant transmission mode for each pixel spectrally red-shifts from 400–to–750 nm as the exposure dose increases, due to the thicker insulator layer, comparing favorably with simulation. The optimization of processing parameters for fabrication is detailed in Supplementary Note 2. Transmission of up to ~75% and relatively narrow FHWMs of ~50 nm are observed in **Fig. 2b** (i), with Δz thickness values up to ~150 nm (**Fig. 2b,ii**) in agreement with the simulation results (**Fig. 2a** and Supplementary Fig 1).

Two different dose-modulated MIM arrays (**Fig. 2c, d**) were then fabricated (i), clearly achieving varying colors (ii) as a result of variations in the cavity height (iii). For isolated (**Fig. 2c**) and dense (**Fig. 2d**) pixel arrays, the EBL proximity effect[28] leads to variation in the final thickness values and hence spectral response for an identical dose range. We therefore determined an empirical correction (decrease) to the dose range in order to achieve the desired spectral response for dense pixel arrays (see Supplementary Note 2.2). For our relatively thin (≤200 nm) MIM-based MSFAs filters, the angular dependency is lower than typical multilayer alternating index filters. For larger chief ray angles of up to 30º, which are typical in smartphone-based image sensors, simulations show a peak wavelength position shift (Δλ) of ~12 nm and ~25 nm for transverse magnetic (TM) and transverse electric (TE) polarized input light, respectively (Supplementary Fig 3).

To demonstrate the versatility of our MIM-based grayscale color approach, we next fabricated a variety of spatially varying optical filter designs, including several different MSFAs, on the same glass chip (**Fig. 3a**). Due to the high sensitivity and high resolution negative tone e-beam resist, the patterning exposure is relatively fast (~several mm$^2$ min$^{-1}$). The MSFAs include two designs at 10 μm pixel pitch: a 2x2 RGB+NIR array (**Fig. 3b**) and an ordered 4x3-band array (**Fig. 3**c). Each array has total dimensions encompassing 2 mm x 4 mm, slightly larger than the active area of the image sensors used for imaging (Supplementary Note 2.2). From a fabrication standpoint, the only difference between the two arrays is the exposure dose of each pixel. MSFAs across the UV-visible-NIR can be easily fabricated (**Fig. 3**d—g) as well as higher order mosaic designs, linear filter arrays to unusual pyramidal structures and spiral phase plates (see Supplementary Fig 12). For conventional commercial techniques to achieve similar design variety and complexity on a single chip with the high optical performance shown here would require many lithographic steps, materials and masks, which is very process intensive and thus both expensive and time consuming.

We then exploited our MIM-based MSFAs for snapshot multispectral imaging. MSFAs containing a Bayer filter (3-band, RGB) and a higher-order (9-band) MSFA were fabricated on the same glass chip. The R, G and B bands of the Bayer MSFA filter (**Fig. 4**a,b) have peak transmission efficiencies of 75, 82 and 88% (±6%), respectively. With center wavelength 640, 546, and 427 nm (±7 nm) and FWHM 55, 66 and 76 nm (±7 nm), respectively.

The Bayer MSFA was then used to image a multispectral test scene (Macbeth ColorChecker with Rubik's cube, **Fig. 4**e). In commercial CMOS image sensors, MSFAs are fixed directly atop the pixels, however, in this proof of concept experiment our MSFAs are fabricated on glass and hence are positioned ~1mm from the pixel surface (*Methods*). The filter pixels are larger than the sensor pixels to accommodate for this. Each filter pixel has dimensions 11 μm x 11 μm, corresponding to a 5 x 5 array of the image sensor pixels. At an image sensor resolution of 1920 x 1080 the 1:5 trade-off in spatial resolution means the effective resolution of our images is 384 x 216. Nonetheless, we have verified fabrication of G-EBL MIM-based pixels with lateral dimensions down to 460 nm (Supplementary Note 2.5) showing the approach is scalable to much smaller pixel sizes suitable for modern CMOS image sensors. With our Bayer filter positioned in front of the monochrome image sensor (**Fig. 4**a) a raw image can be captured and demosaiced (Supplementary Fig 17) to form a standard 3-band RGB image (**Fig. 4**e). Without enhancement (i.e. standard color correction techniques used in imaging) the captured raw image provides accurate reconstruction of the multispectral test scene when compared to a commercial RGB color camera (**Fig. 4**e,ii).

Finally, we characterized the custom 9-band (8-band + 1-reference band) MSFA (**Fig. 5**a,b) and placed it in front of the image sensor (**Fig. 5**c) to demonstrate snapshot multispectral imaging. From shortest to longest wavelength, the 8-band array has center wavelengths of 415, 463, 518, 572, 621, 660, 706, 725 nm (±6 nm), FWHMs of 81, 75, 66, 58, 55, 53, 48, 46, 79 nm (±7 nm) and peak transmission efficiencies of 69, 75, 74, 72, 69, 68, 65, 59% (±7%), respectively. The array contains a single reference band (dark pixel) that is not exposed during lithography so consists of two Ag mirror layers without a cavity. As both the Bayer and 9-band MSFAs are fabricated on the same chip (akin to **Fig. 3**a) a simple vertical positional translation enables us to switch between standard RGB and multispectral imaging modes. Initially we used a supercontinuum white light laser source as a collimated input, recording the intensity response through the MSFA at 10 nm steps with center wavelengths in the range 450—750 nm (see Supplementary Video 1). The MSFA can be visually identified (**Fig. 5**d) and the correspondence between input wavelength and intensity distribution across the pixels is evident (see also Supplementary Video 1). Multispectral imaging was then performed using a dynamic test scene consisting of spatially separated optical bandpass filters backlit with a white LED floodlight (**Fig. 5**e,i). The raw color image (**Fig. 5**e,ii) is composed of the monochrome intensity multiplied by the 9-band MSFA matrix, showing the power distribution across the 9-bands. The demosaiced images (see Supplementary Fig 17) for four bands of the MSFA (**Fig. 5**e,iii) show the ability to clearly discriminate the spectral information within the test scene.

Our bespoke 9-band MIM-based MSFA was able to outperform alternative approaches for color filter fabrication, such as plasmonic and high-index dielectric nanostructure arrays/metasurfaces. For example, in comparison to recent progress in spatially variant spectral filters[8,10–18], our MSFA transmission bands with (average transmission 70±5% and average FWHM 60±11%) are generally narrower, have higher transmission efficiencies, exhibit no polarization dependency (up to high angle of incidence chief ray angles) and have no additional orders within the visible-to-NIR wavebands. Snapshot multispectral imaging was undertaken with both the Bayer 3-band and custom 9-band MSFAs, which when demosaiced showed promising performance when applied to their respective test scenes.

A key challenge for MSFA fabrication to achieve widespread adoption is scalability and compatibility with standard micro/nanofabrication processing. The framework presented here uses G-EBL to produce low-volume bespoke designs. As G-EBL is a direct write system, with limited throughput, it is typically incompatible with large parallelized processing techniques.

Nevertheless, the concept itself can be easily translated to photolithographic techniques/ larger scale (see Supplementary Fig 18), with conventional masked or mask-less photolithography, utilizing either: grayscale amplitude photomasks (i.e. tailoring the thickness of the chromium regions to change opacity); binary amplitude photomasks in conjunction with flood exposure energy modulation; or digital-mirror-device (DMD) based maskless grayscale lithography.[29] Moreover, translation to photolithography would mitigate problematic aspects of this work based on EBL, such as reducing proximity effects (due to electron scattering events), improving reproducibility and hence consistency of the filter pixels' optical properties. Materials compatibility for CMOS manufacture /processing is discussed in Supplementary Note 2.6.

In summary, we have presented a unique approach for producing high efficiency (~75%), narrowband (~50 nm) highly customizable MSFAs operating across the visible to NIR using a single lithographic step. G-EBL is utilized to generate customizable insulator thickness profiles in MIM geometries producing optical filters spanning the UV–visible–NIR. We demonstrate application to imaging by placing the fabricated MSFAS in front of a commercial image sensor and performing pixel-wise discrimination of different wavelength bands in a multispectral test scene. Current manufacturing methods for producing spatially variant optical filters, from linearly variable filter arrays to MSFAs, will typically use N-lithographic steps for N-wavelength bands and require a variety of materials,[7,30] limiting their customizability. By contrast, this versatile approach requires only a single lithographic step and the same materials for each band, making it highly customizable. Further, unlike other single step approaches with a single material layer such as nanostructured filters (plasmonic, all-dielectric or otherwise),[10,19,20] it does not require ultra-high resolution patterning and is relatively polarization-independent. MIM-based MSFAs could therefore enable a whole new range of bespoke, low-volume multispectral image sensors targeted to different applications. Further, we envisage the novel methodology reported here could play a significant role in reducing the fabrication complexity and cost of intricate optical filter devices for high-volume MSFAs such as the Bayer filter.

# Methods

## *Fabrication Techniques*

MaN-2400 series negative tone photoresist [*Micro resist technology GmbH*] is utilized for this study due to its high-resolution capability for EBL in combination with relatively high sensitivity. Double-side polished borosilicate (Borofloat 33) glass [*Pi-kem*], thickness 500±25 µm, is diced to 1 cm$^2$ samples. The glass samples are cleaned in successive ultrasonic baths of acetone [*Fisher Scientific*] and isopropyl-alcohol (IPA) [*Fisher Scientific*] for 10 min, blow-dried with ultra-high purity (UHP) compressed N$_2$ and dehydrated at 200ºC for 10 min.

A 1.5 nm Ti adhesion layer is thermally evaporated [*Edwards E306 Evaporator*] (base pressure ~2x10$^{-6}$ mbar, deposition at 0.1 nm. s$^{-1}$), followed by a 26 nm layer of Ag (with relatively fast deposition, 0.2–0.3 nm. s$^{-1}$, for improved optical performance), followed by a second 1.5 nm Ti layer. The first Ti layer promotes adhesion between the glass and Ag, the second increases the wettability of Ag for resist spin-coating and increases chemical stability by reducing Ag oxidation. The optimal thickness of the Ag is determined through simulations (Supplementary Fig 1) trading transmittance against FWHM. The thickness of the Ti layers is such that resist wettability is increased and adhesion is promoted with minimal effect on optical transmittance. MaN-2405 eB resist is spin-coated on top of the samples at 5,200 rpm for 45s to form a ~350 nm layer, then baked at 90ºC for 3 min. High voltage (80 kV), high current (4.2 nA) EBL (nB1, *Nanobeam Ltd.*) is used for the patterning. The bottom metallic mirror layer additionally acts to dissipate accumulated charge during electron beam exposure. The MSFAs have total area dimensions ~1.1x greater than the image sensor area (4.85mm diagonal) to correct for the proximity effect (Supplementary Fig 8) and ensure all sensor pixels are utilized. The effect of stitching error is reduced due to the rectangular geometry (edges) of the patterns corresponding to the main-field and sub-field fractures. No sample registration marks are used for the samples shown in this study. The high current, in combination with low critical dose (due to inherent high sensitivity) of the resist, allows for fabrication over relatively large areas (~millimeters) in quick time periods. The critical parameters in this grayscale-to-color study are the exposure dose and development conditions, which are determined empirically through a variety of dose tests (Supplementary Note 2.1). For this study, a dose range 5–75 µC cm$^{-2}$ is used and full concentration AZ-726-MIF [*AZ Electronic Materials*] developer solution for ~10s, followed by two DI water (stopper) rinses for 4 min and UHP compressed N$_2$ blow dry. A post-development bake (90 ºC for 30s)—in which the resist is brought within close proximity to its glass transition temperature—is subsequently performed which yields a smoother surface before the second mirror deposition and improves optical performance (Supplementary Note 2.3). The top-metal, a 26 nm layer of Ag, is thermally evaporated (deposition at 0.2–0.3 nm. s$^{-1}$) followed by a 12 nm layer of MgF$_2$. This final encapsulation layer adds chemical and mechanical stability to the MSFAs with minimal, if not improved, effect on optical properties (Supplementary Fig 2).

## Optical and Morphological Characterizations

Surface morphology is characterized using an Atomic Force Microscope (AFM) [*Asylum Research MFP-3D*] in conjunction with Al-reflex-coated Si probes [*Budget Sensors, Sigma Aldrich*] operated primarily in tapping mode. Scan speed, voltage set-point, and drive amplitude are modified dependent on the feature morphology. *Gwyddion* software is used for the AFM data visualization and analysis. The raw surface data is plane levelled, scars (strokes) and noise minimized, and the resultant data is presented in 3D topography form. The average height (and standard deviation) of each pixel (such as in **Fig. 2**b, ii) is obtained using the in-built statistical analysis toolbox. A *LEO Gemini 1530VP* field emission scanning electron microscope (SEM) operating at 1–5 keV is used for imaging the surface of samples (In-lens operation), which are fixed on angled SEM stubs for non-normal incidence imaging. *Carl Zeiss* software [*SmartSEM*] is used to control the SEM and obtain images at several magnifications. The optical characterization is performed using a modified *Olympus BX-51* polarizing optical microscope (Halogen light source with IR filters removed) attached via a 300 μm core multi-mode optical fiber [*Ocean Optics* OP400-2-SR MMF] to a UV–visible spectrometer [*Ocean Optics HR2000+*] and second optical arm to a digital camera [Lumenera Infinity-2 2MP CCD] for surface imaging (Supplementary Fig 15). The spectra are normalized to transmission through equivalent thickness borosilicate glass (bright state) and no input light (dark state) using *Ocean Optics* OceanView software.

Further characterization is performed using a supercontinuum white light source and tunable filter [*NKT Photonics: SuperK COMPACT source and SuperK VARIA filter*]. Supplementary Video 1 shows the normalized intensity map of the image sensor as a function of increasing input wavelength (450—750 nm; 10 nm linewidth), generated with a supercontinuum white light laser source. The laser is fiber coupled, expanded and collimated to be used as an input source to the MSFA-sensor. There are no imaging optics in the system. The video shows (*right*) how the intensity response changes as a function of input wavelength, with the respective MSFA (*left*). The spatial progression of the highest intensity regions follows that which is expected of the MSFA array's spectral characteristics.

For the imaging experiments, the test scene is composed of a *Macbeth ColorChecker* chart (A5 size) along with a Rubik's cube, which is imaged via a series of lenses through the custom MSFAs onto a CMOS image sensor (Supplementary Fig 16). A USB 3.0 monochrome 2MP *Basler daA1920-30um* area-scan camera is used [*Aptina MT9P031 CMOS image sensor*], with a total sensor size of 4.2 mm x 2.4 mm, resolution of 1920 x 1080 and 2.2 μm x 2.2 μm pixel size. The image sensor is mounted at the end of a custom optical cage-system using a 3D printed [*Ultimaker 2+*] mount. An in-house built XYZ translation mount holds the MSFAs, which are fabricated on 10 mm x 10 mm borosilicate glass chips. The imaging optics consist of three achromatic AR-coated lenses (*Thorlabs LSBO8-A series*): the first (a concave lens) de-magnifies the scene, the second collimates this virtual image (placed at the focal point of the first lens) and the third focuses the light onto the image sensor, through the MSFA mounted in front of it. An aperture stop is located after the third lens, limiting the range of ray angles impingement on the MSFA and thus onto the image sensor. The MSFAs, fixed in a custom 3D printed mount, are brought close to the borosilicate cover glass (thickness 0.4 mm) of the image sensor. Using the image sensor manual (*Micron* MT9P031 manual and *Basler* AW001305 documents) to determine the physical sensor geometry, the minimum distance of the MSFA from the image sensor die (plane) is estimated at ~0.525 $\pm$ 0.05 mm. The MSFA is translated in XYZ in order to align the pixels of the filter array with the pixels of the image sensor. For the MSFA imaging results, a series of optical bandpass filters (*Thorlabs FKB-VIS-10 series*; 10 nm FWHM) are utilized in a filter wheel mount, backlit with 50W white light (4000K) floodlight LED array. The reflected light from the object test scene is imaged through the MSFA onto a monochrome image sensor. The process of starting with the raw 2D intensity matrix (from the image sensor), which has no wavelength specific information, and determining the wavelength specific information (N-band x 2D-image data) is performed using custom code in MATLAB (Supplementary Fig 17).

## *Numerical Simulations*

A commercial-grade simulator (*Lumerical FDTD solutions*) based on the finite-difference time-domain (FDTD) method was used to perform the calculations.[27] MIM stacks are simulated using a dielectric between two metal layers (z-dimension). Periodic boundary conditions are used (x-y boundaries of the unit cell) and perfectly matched layers (z-boundary) along the direction of propagation. A uniform 2D mesh (Yee-cell) with dimensions ≤1 nm and broadband-pulse plane-wave (350–1000 nm) injection source at a significant distance (several microns) above the sample are used. For the E-and-H-field intensity plots, an additional finer mesh is included, whereby the smallest cubic mesh size is <0.01 nm (z-direction). Complex dispersive material models are used for Ag (Johnson and Christy model) and $SiO_2$ (material data), whereas a real-only refractive index of 1.65 is used for MaN-2400 series photoresist (*Microchem: Material data sheet*) and 1.4 for the transparent $MgF_2$ capping layer. Transmittance and reflectance values are calculated from 1D power monitors positioned above the range of structures and source injection.

# Figures

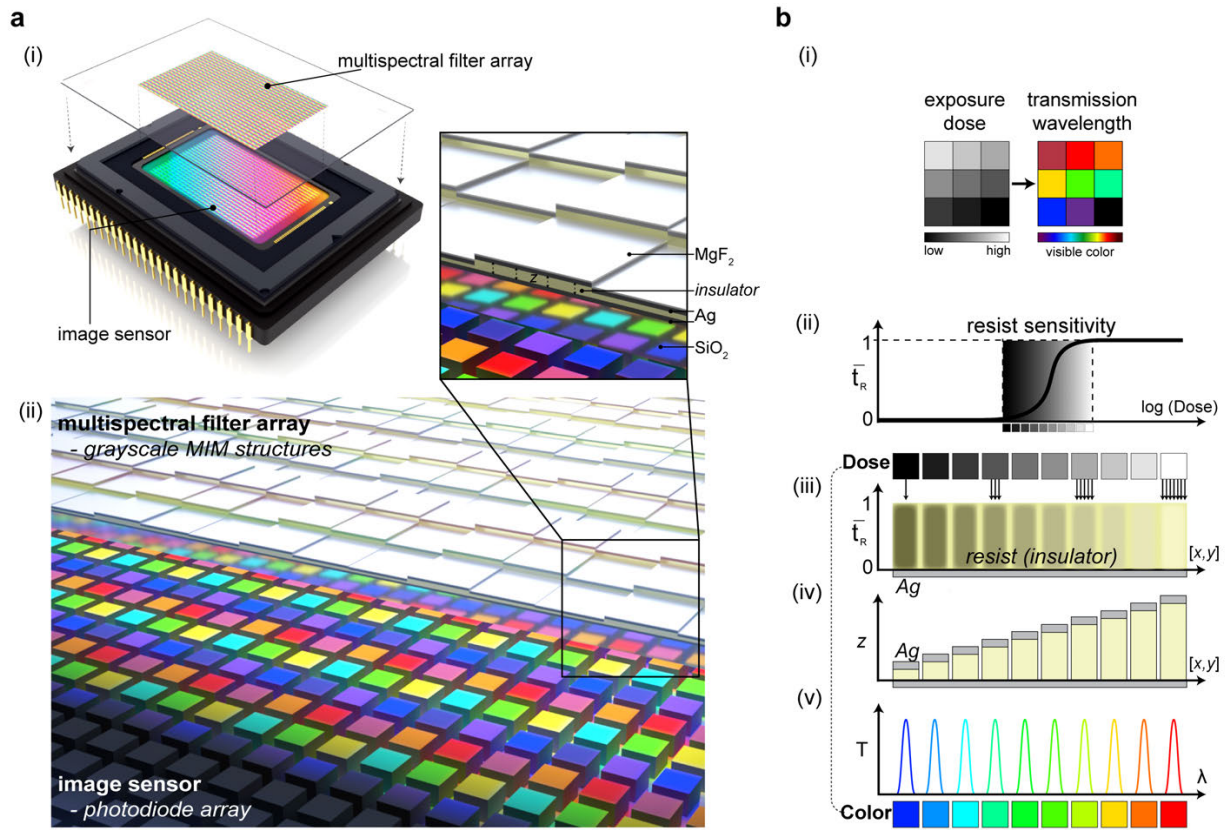

**Fig. 1**

**Multispectral filter arrays (MSFAs) using grayscale lithography with metal-insulator-metal geometry.**

**a**, *Schematic*: (i) Using a customized MSFA atop a monochrome image sensor for multispectral imaging. (ii) 3D MIM structure of MSFA with *inset* detailing layers. The wavelength transmitted to each pixel below the MIM structures is controlled with the single-step lithographic fabrication process. **b**, MSFA fabrication process: (i) a spatially varying grayscale exposure dose results in a spatially varying wavelength transmission profile. (ii) Calculated grayscale exposure dose profile corresponding to remaining resist thickness profiles ('resist sensitivity' curve). An ultrathin noble metal (Ag) layer on glass ($SiO_2$) acts both to dissipate accumulated charge and as the bottom-mirror of the filter. (iii) A spatially variant dose modulated exposure leaves a 3D resist profile post-development. (iv) Post-metal-deposition: with a top metal (mirror) layer, the spatially varying 3D resist profile acts to filter the light according to the eigenmode solution of the stack. (v) Final spectral transmittance profiles of MIM structures.

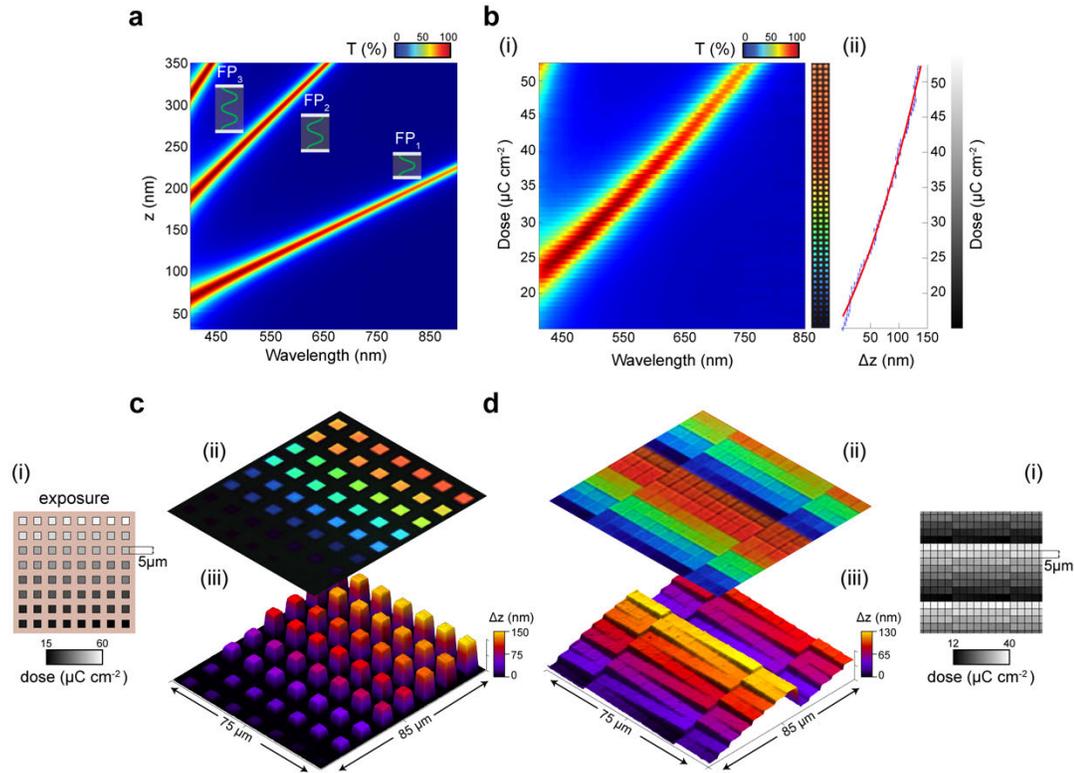

**Fig. 2**

**Grayscale exposure dose to color: experimental verification**

**a**, Finite difference time domain (FDTD) simulations of the optical transmission from a continuous Ag-based MIM cavity as a function of varying insulator thickness, with geometry: $SiO_2$(bulk)–Ag(26nm)–Resist (n=1.653)–Ag(26nm)–$MgF_2$(12nm). **b**, Experimental demonstration of grayscale-to-dose pattern with the same layers as in **a**: (i) Transmission spectra from dose modulated 5μm x 5μm squares (optical micrograph shown in inset), which results in increasing thickness and hence varying peak wavelengths, (ii) measured curve linking dose and thickness. Only the first-order resonance is present at doses, but for higher doses (>50 μC cm-2), the second-order mode is also excited. **c**, Dose-modulated 5μm x 5μm pixel array with 10μm spacing: (i) dose-modulated pattern, (ii) optical micrograph, and (iii) corresponding AFM data. **d** Same as **c** but with zero dead space.

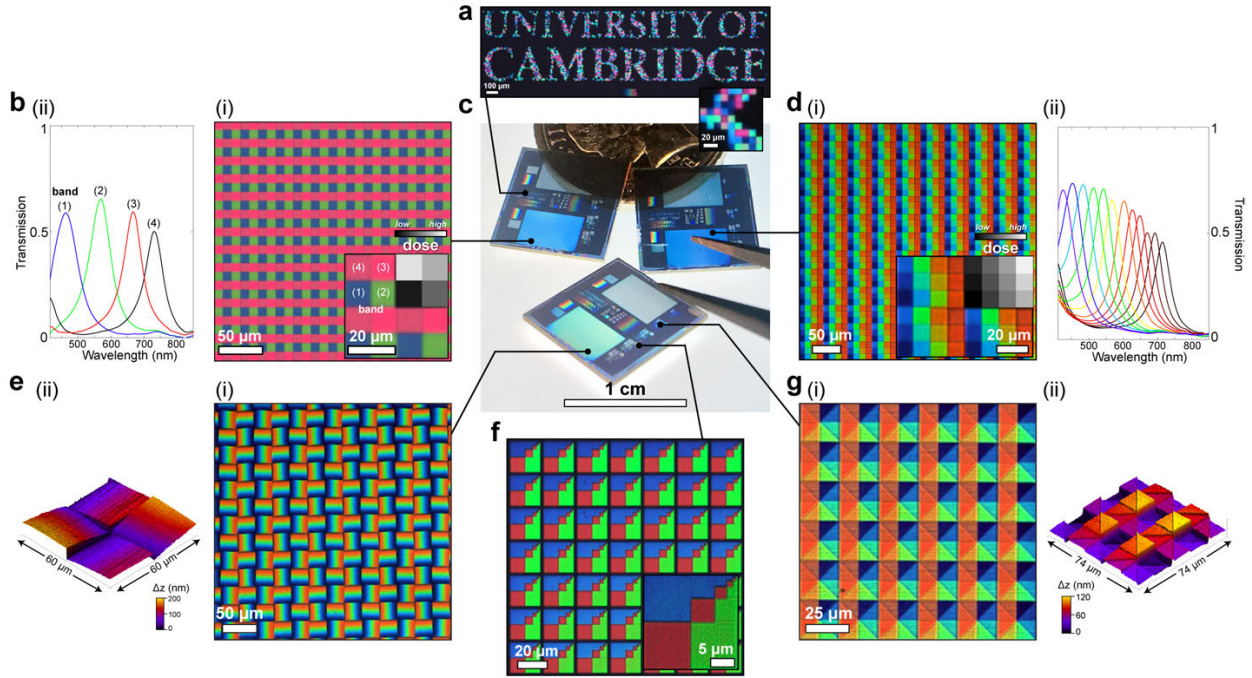

**Fig. 3**

**Demonstration of the versatility of MIM-based MSFAs through patterned design variety**

**c**, Photograph of three identically processed chips with a range of patterned designs on each chip with varying complexity. Each chip is processed in a single lithographic step in G-EBL. **a**, Optical micrograph (with magnified inset) of the University of Cambridge logo text composed of 10 µm pixels with a randomized exposure dose profile, hence random colors in transmission. **b**, RGB+NIR MSFA (bands labelled in inset) with (i) optical micrograph in transmission, and (ii) respective transmission spectra of the wavelength bands. **d**, spectrally 'ordered' 4x3 mosaic, (i) optical micrograph, (ii) transmission spectra. **e**, 25 µm linearly variable filter pixel design, (i) optical micrograph, (ii) AFM micrograph of the unit cell showing the in-plane height variation. **f**, Optical micrograph (with magnified inset) of an array of RGB pixels with exponentially ($2^{-n}$) decreasing pixel width, starting from 10 µm. **g**, 25 µm discrete spiral phase pixel design (i) optical micrograph, (ii) AFM micrograph. Transmission spectra represent averages of 5- different acquisitions, taken at random positions across the array.

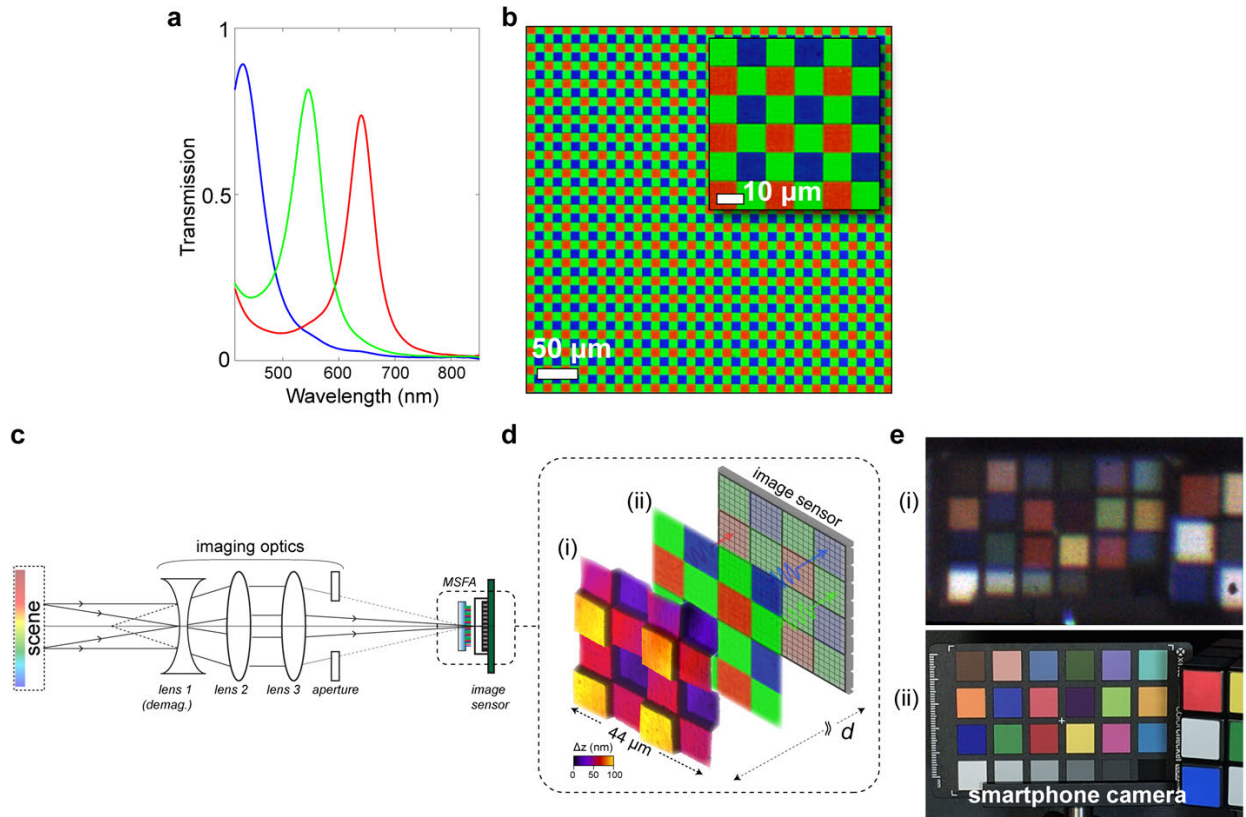

**Fig. 4**

**Imaging through a customized Bayer (RGB) MSFA.**

**a**, Transmission spectra of the 3-bands (e.g. RGB) of the Bayer filter, with **b**, optical micrograph of the mosaic. **c**, Schematic of the optical imaging setup (see Supplementary Note 5), with **d**, physical representation of the MSFA in front of the image sensor: (i) experimental AFM micrograph, (ii) optical micrograph and image sensor schematic, where *d* is the distance of the MSFA from the sensor plane (~1 mm). **e**, A snapshot of the imaging test scene, including Macbeth ColorChecker chart and Rubik's cube, captured with a monochrome image sensor through our filter (ii) and using a conventional smartphone (ii) for reference. In (i), aside from demosaicing, there is no post-processing (enhancement) of the color.

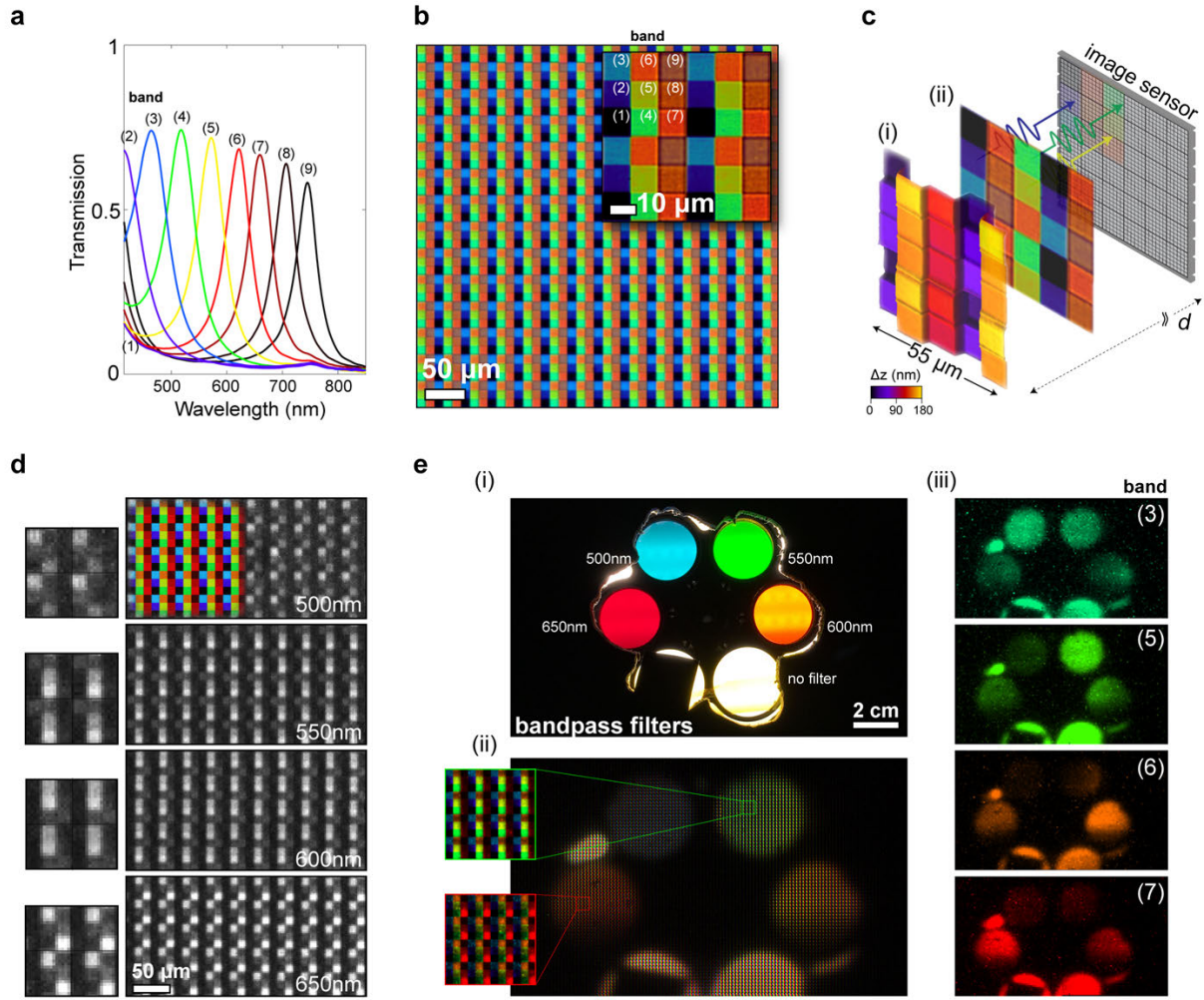

**Fig. 5**

**Multispectral imaging through the custom 3x3 MSFAs.**

**a**, Transmission spectra of the 9-band (labelled) MSFA, **b**, optical micrograph of the mosaic. **c**, Schematic representation of the MSFA in front of the image sensor: (i) Experimental AFM micrograph, (ii) optical micrograph and image sensor schematic, where *d* is the distance of the MSFA from the sensor plane (~1 mm). **d**, 2D intensity matrices from the monochrome image sensor (captured through the MSFA) with illumination from a supercontinuum source (Supplementary Note 3.3 and Supplementary Video 1) at four different center wavelengths: 500, 550, 600, 650 nm; FWHM 10±2 nm. **e**, *Multispectral imaging*: (i) Reference image of test scene (comprising 4 bandpass filters across the visible spectrum) taken using a conventional smartphone image sensor. (ii) The equivalent 'raw' color image captured using the monochrome image sensor through our MSFA with individual MSFA pixels visible. (iii) Demosaiced color-coded images for four different wavelength bands obtained from the 'raw' image in **e** (ii), with labels denoting band.

# Data availability

The data that supports the findings of this work will be made available online [open access] upon publication.

# Acknowledgments

This work was supported by a CRUK Pioneer Award (C55962/A24669) and other CRUK awards (C47594/A24669, C14303/A17197, C47594/A16267). C.W. would like to acknowledge the support of Wolfson College, Cambridge. The authors would like to acknowledge the following researchers for their fruitful discussion and help with the work: Catherine Fitzpatrick, Girish Rughoobur, Dale Waterhouse, Jonghee Yoon, Siri Luthman, Ventsislav Valev and Christian Kuppe.

# Authors' contributions

C.W. conceived the idea. C.W. carried out the design and simulation of the devices. C.W. fabricated the devices and performed the experimental measurements. C.W and S.G characterized the devices. G.S.D.G., T.D.W and S.E.B. supervised the study. C.W, S.G. G.S.D.G, T.D.W and S.E.B analyzed the data and wrote the manuscript.

# Competing interests

The authors declare no competing interests.

# SUPPLEMENTARY INFORMATION

## Grayscale-to-color: Single-step fabrication of bespoke multispectral filter arrays

## Supplementary Notes

### 1. Simulations

### 2. Fabrication

## Figures

## References

# Supplementary Notes

# 1. Simulations

## 1.1 Effect of insulator (resist) thickness on transmission

**Supplementary Figure** 1 presents simulation results (Lumerical FDTD solutions[1]) of the optical transmission through a metal-insulator-metal (MIM) stack. Due to its simplicity and customizability, the FDTD method used here is chosen as opposed to a more analytical transfer matrix / rigorous coupled wave analysis (RCWA) method.[2] The simulations are used to confirm: the origin of the resonant modes; resist (insulator) thickness required to achieve operation in the visible, and; how different parameters (e.g. layer thicknesses) affect the transmission characteristics. Hence we use the simulations to determine the required layer thicknesses of the multispectral filter arrays (MSFAs). The simulations for the MIM stacks (in *Supplementary Figure 1: main manuscript*) were performed using periodic boundary conditions parallel to the propagation direction of the incident electromagnetic wave and perfectly matched layers tangential to the incident wave. Plane waves (300—1000 nm spectrum) are injected toward the structure from the $SiO_2$ side and a power monitor several microns away from the structure is used to calculate the transmission. The resist (insulator) layer is assigned a real-only refractive index value (n = 1.635), obtained through the MaN-2400 series photoresist datasheet [*Micro resist technology GmbH*]. The other layers are assigned a complex and dispersive refractive index from Lumerical FDTD solutions: Ag model - Johnson & Christy; $SiO_2$ – Palik; $MgF_2$ – model fitted to tabulated values. Ag is chosen as the mirror material due to considerations of its complex dielectric function in the visible and near-infrared part of the spectrum.[3–6] Based on **Supplementary Figure** 1, the simulations indicate that as the insulator layer increases the Fabry-Perot-type mode red-shifts accordingly (due to an optical path length increase) and the mirror thicknesses (c) control the Q-factor of the cavity (linewidth). As a result, a metal thickness of ~25—30 nm results in a suitable tradeoff between transmission and FWHM.

## 1.2 Addition of an encapsulation layer

Depositing an inert capping layer on top of the MIM stack provides chemical and mechanical durability, preventing oxidation and increasing rigidity. A material such as $MgF_2$ (similar to quartz) provides these qualities. It is relatively inert, mechanically rigid, optically transparent and moreover, relatively straightforward to deposit post-metallization of the second mirror. **Supplementary Figure** 2 shows the simulation of the transmission of the Ag (26 nm)-MIM stack (125 nm insulator) as a function of $MgF_2$ encapsulation layer thickness (0—50 nm). A dispersive material model[7] is used for its refractive index. It is observed the transmission peak slightly shifts to longer wavelengths with increasing $MgF_2$ thickness and increases a small amount. The FWHM also gradually increases. We therefore conclude that the encapsulation layer provides negligible degradation in transmission characteristics, and if anything, slightly improves them.

## 1.3   Angle dependency

In imaging optics, a high *F*-number (numerical aperture ~ 0) typically implies parallel rays incident on the sensor array, while a low *F*-number implies the rays arrive at an angle (numerical aperture > 0). For multi-layered MSFAs, the spectral response is often a function of incidence angle, which in combination with polarization angle, affects the transmission characteristics of the filter.

**Figure 3** shows FDTD simulations (Bloch boundary conditions and angled source using BFAST conditions) of the MIM structure under orthogonal incident polarizations (TE and TM) for varying angles of incidence up to 60º. For conventional CMOS image sensors, the chief ray angle (CRA) – broadly defining the cone of angles incident upon the center top of the pixel – can be up to ± 30° (with the larger angles, e.g. 30°, more commonplace in smartphone sensors). It can be observed that, between 0 — 30º incident angle, the peak spectral position varies from ~590—578 nm (Δλ ~ 12 nm) for TM and from ~590 — 565 nm (Δλ ~ 25 nm) for TE. The transmission intensity remains relatively constant (~85%) across these angles. The FWHM (~36 nm at 0º, for TM and TE) narrows slightly by ~5 nm for TE, and for TM widens by ~15 nm (at 30º). Beyond 30º, the spectral shift increases more significantly, especially for TE-polarization.

For interference filters in general, there is typically a blue-shift of the resonant peak arising from a phase-shift reduction in the dielectric layer for larger angles.[8] For interference based optical filters, there are methods to compensate for such a spectral shift, including: in-plane nanostructuring; incorporation of additional dielectric layers; and addition of microlens arrays.[8,9] Conversely, for the relatively thin (<200 nm thick) Ag-MIM filters here, that only support first-order FP-type modes, the angular dependency is somewhat reduced compared to thicker structures with a high number of alternating index layers. The small vertical-to-lateral aspect ratio of the MIM pixels result in relatively small angle dependency.

# 2. Fabrication

## 2.1   Dose variation

As described in the main manuscript, the effect of exposure dose and correct choice of development time (and developer) controls the final thickness of the remaining resist (insulator) in a MIM cavity, hence controlling the center position of the transmission spectra. To demonstrate this, **Figure 4** shows the transmission spectra of a set of 5 µm pixels which vary in exposure dose across three different development times. It can be observed—both quantitatively in (a) and visually in (b)—that for a constant dose range (0.1–0.7 Cm$^{-2}$ here) the position of the peak blue-shifts with increasing development time. As the developer is selectively removing resist that has not been sufficiently cross-linked (due to MaN-series photoresist being negative tone), a longer development time results in more resist being removed, hence thinner cavity and shorter wavelength mode. This is further illustrated in **Figure 5**, which shows a rectangular array with transmission wavelength across the visible spectrum and respective SEM micrograph.

**Figure 6** shows a custom dose/resolution pattern fabricated 4 times (i—iv) across a sample, with the dose incremented by +0.125 Cm$^{-2}$ each time. The pattern itself consists of 100 separate layers; each layer corresponds to a linearly increasing dose within the range 0.1–0.4 Cm$^{-2}$. The figure, and in particular the magnified regions in (b) and (c), visually shows the main operating principle of this work: exposure dose controls the wavelength of transmission.

## 2.2   Proximity effect

In EBL, the proximity effect is the unwanted exposure of regions adjacent to the pattern being exposed due to electron scattering events in the resist. The proximity effect can be lessened through the translation of the grayscale MSFA approach to larger batch processing i.e. photolithography. However, for this work (in which EBL is utilized) as the density of features increases, the proximity effect is more pronounced. In this work, each filter pixel has its center wavelength defined by a specific exposure dose. As a result of the proximity effect, the total dose applied to a specific region (pixel) is additionally dependent on the dose applied to surrounding pixels.[10,11]

The proximity effect can be observed by comparing the patterning of isolated pixels (i.e. arrays with non-exposed spacing between pixels) to dense arrays; the dose required to achieve a specific wavelength (resist thickness) is lower in dense arrays than it is in isolated regions. **Figure 7** (a) shows an optical micrograph (transmission) of the dose test array, in which the regions (1) and (2) are arrays of equally sized pixels which both also equally increase in exposure dose (from ~0.17—0.52 Cm$^{-2}$; 10s development time), but with 'isolated' and 'dense' configurations respectively. It is observed that the arrays in (2) are highly red shifted in transmission indicating a larger thickness in remaining resist and thus greater accumulated exposure dose. This is due to the unwanted cumulative adjacent exposure from the neighboring pixels. **Figure 7** (b) is an additional example of the effect: a 1951 USAF resolution target, in which each element of the line triplets is given a different exposure dose. The final thickness/filtered wavelength is a function of spatial position within the rectangle as the averaged dose density is larger at the center of the rectangle than it is in the corner/edges.

The impact of the proximity effect in a Bayer filter array was investigated by examining the transmission characteristics of a 3-channel RGB array as a function of position from the edge of the array **(Figure 8).** The center wavelength of the transmission peak (in both green and blue pixels) appears to remain relatively constant following a sharp change approximately 50–100 μm from the edges. This is likely due to the cumulative dose density remaining approximately constant for pixels away from the edge of the array, a consequence of the periodic array pattern. A simple empirical correction adopted for this work was to 'over pattern' each MSFA, such that the area of interest (image sensor area) is >100 μm from the edge of the MSFA pattern. This approach also demands reducing the dose profile to compensate for increased cumulative exposure in the central region. It would also be possible to perform Monte Carlo electron scattering simulations for each pattern to optimize the dose patterns and avoid this empirical correction, however, the commercial software to perform these simulations was not available for this study.

## 2.3   Resist thermal reflow

Thermal reflow is a fabrication processing technique that involves the thermal treatment of a photoresist (post-development) such that the resist is brought to a temperature $\gtrsim$ glass transition temperature.[12] By doing so, the resist 'reflows', fully or partially depending on the temperature and time, which can be used to smooth the resist. The technique, for example, can be used to turn staircase-like 3D-pattens to 3D slopes,[12] or to fabricate microlens arrays. In this work, shown with several examples in **Figure 9**, we used thermal reflow to smooth the resist surface post-development, but pre-second metal mirror deposition, to flatten/smooth the second mirror surface, narrowing the FWHM and boosting transmission efficiency.

## 2.4   Variability in optical performance across array

The intra- and inter-chip variability of the optical characteristics of fabricated MSFAs is shown in **Figure 10** and **Figure 11**. For each MSFA, a range of unit cells were chosen at random across the array (but at least 100μm from the border of the array due to the proximity effect issue described in Section 2.2) and the filter spectra were recorded and analyzed using an optical microscope. **Figure 10** shows the variability in peak wavelength across different pixels for RGB MSFAs (i.e. Bayer mosaics) for three different processing recipes corresponding to three differently processed separate chips (listed below, *Recipes 1—3*). (a) and (b) are of two different dose (*D*) profiles for the 3-channels. For each 'recipe', spectra of four randomly positioned unit cells were analyzed (4x 3-channels = 12 points) i.e. 4 transmission spectra per channel (wavelength band).

**Figure 11** shows box plots of the optical transmission characteristics of a range of MSFA geometries fabricated across three different chips (i.e. the three different recipes from **Figure 10**). These include 2 x 3-channel designs (RGB1 and RGB2), RGB+1, and 3x different 3x3 mosaics (each with a varying dose profile range). The three different recipes correspond to the following processing conditions:

*Recipe 1* = pre-development thermal treatment (90ºC, 60s) + normal processing*;
*Recipe 2* = normal processing*;
*Recipe 3* = normal processing* + post-development thermal treatment (100ºC, 30s).
* *normal processing recipe described in Methods - main manuscript.*

It can be observed from both figures that the variation in optical performance characteristics is minimal within each respective array. For example, the respective channel peak wavelength shift is typically ≲5nm across the arrays and different recipes (**Figure 11b**). Moreover, it can be concluded from these results that the addition of baking steps to the standard protocol enhances the peak transmission. As shown in **Figure 11 (a)** adding a post-development bake (Recipe 3) increases the peak-transmission up to ~80%. The FWHM is also improved (**Figure 11c**) through adding additional thermal treatment; decreasing to ~50 nm in comparison to the standard protocol.

## 2.5    Pixel resolution dose tests

In the main manuscript, we use 11µm x 11µm pixel dimensions, primarily due to limitations with the experimental image sensor setup, however, these length scales can be easily reduced. To demonstrate such scalabilty, we fabricated arrays where exposure dose is varied linearly, with pixel sizes range from: 5.5 µm down to 460 nm (**Figure 13**) Note, 460 nm is not the upmost limit to resolution but for this pixel dimension, the range of lateral-to-vertical aspect ratios of ~9:1 (e.g. ~450 nm:50 nm) to ~9:4 (e.g. ~ 450 nm:200 nm) meaning that they exhibit low aspect ratios and are mechanically stable, hence we suspect the resolution can extend beyond what is demonstrated here. In addition, **Figure 14** shows a series of angled SEM micrographs of 1 µm and 500 nm pixels showing the surface morphology, and their uniformity at these size scales.

## 2.6    Materials considerations for CMOS processing

The properties of materials also present limitations of the current approach for real world applications. For example, the insulator layer. Here we are using a resist (polymer) with a glass transition temperature ≲100ºC. With critical CMOS image sensor manufacture/processing steps requiring greater temperatures (such as wire bonding and wafer level packaging) a more durable and thermally stable insulator layer may be required. Nonetheless, there are many negative-tone resists designed to exhibit high thermal stability (≲250ºC), which can be implemented in this process. Alternatively, a longer-term solution would be to use glass (SiO2), which could be achieved using the grayscale resist as an etching mask for a reactive ion etching step (see Supplementary Fig 19). The use of thin-film Ag similarly provides a potential challenge in the form of long term chemical stability. By encapsulating Ag with chemically inert and optically transparent thin films, such as silicon oxide, this issue may be lessened. Furthermore, a more comprehensive approach would be to replace the Ag with all-dielectric longpass filters(reflectors) operating with a cut-off wavelength of ~400 nm, thus passing the visible-NIR.

# Figures

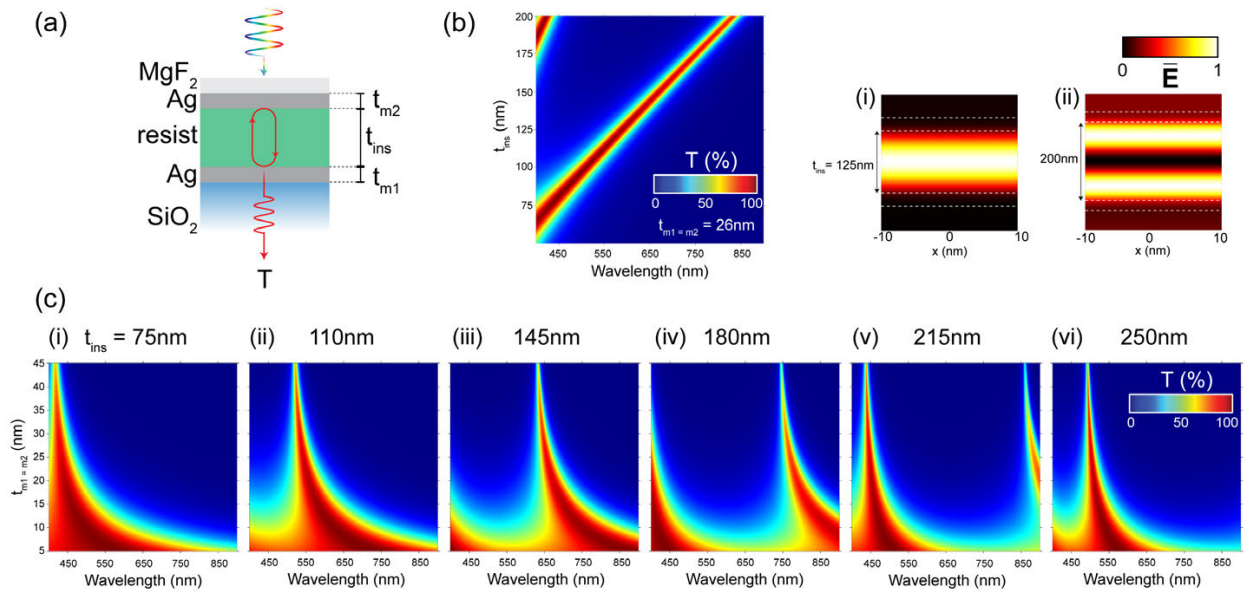

**Supplementary Figure 1. FDTD simulation of the optical transmission from a Ag-MIM stack.** (a) Schematic of MIM geometry used in simulation. (b) FDTD simulation of the transmission from the Ag-MIM-MgF$_2$ (encapsulation) stack as a function of insulator (resist) thickness. (i) normalized electric field intensity map for insulator thicknesses of 125 nm (at λ = 580 nm) and (ii) 200 nm (at λ = 580 nm) corresponding to first-order and second-order resonances respectively. (c) shows the effect of metal mirror (Ag) thickness on the transmission spectra for varying insulator thicknesses (75 – 250 nm).

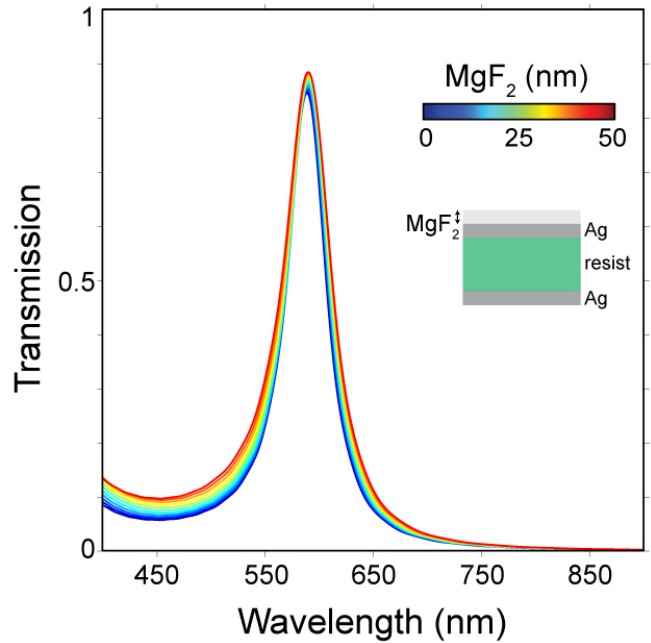

**Supplementary Figure 2. Effect of the addition of a MgF₂ encapsulation layer.** FDTD simulation of the transmission through the Ag-MIM stack in **Figure 1**, but with the addition of an encapsulation layer with varying thicknesses, from 0—50 nm.

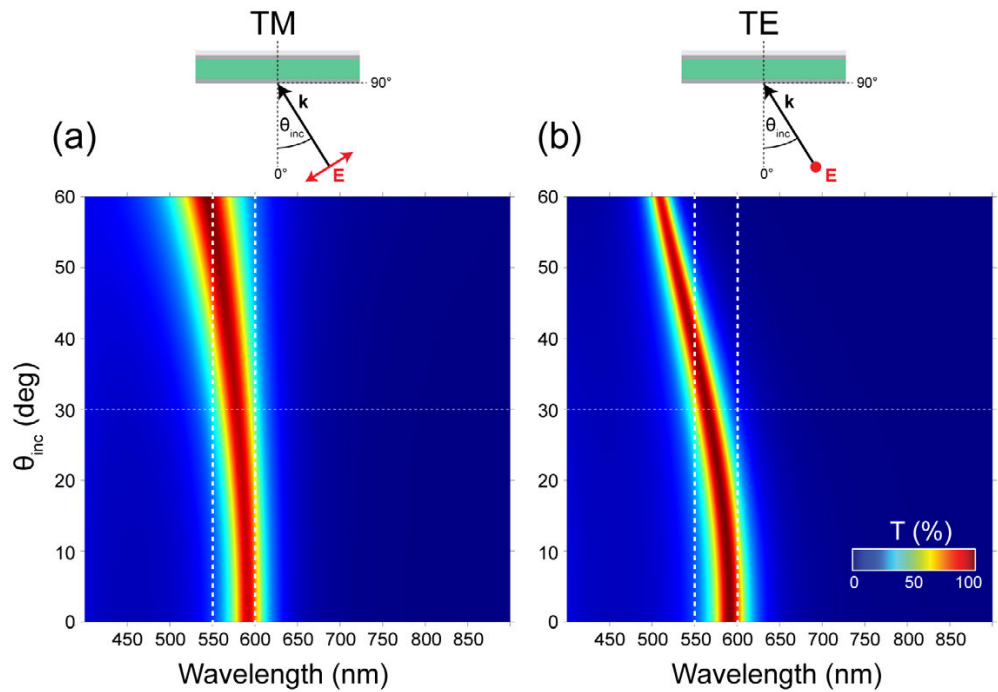

**Supplementary Figure 3. MIM stack transmission spectra.** (a) TM-polarized input wave, and (b) TE-polarized input wave, for various angles of incidence, from 0—45 degrees from normal. The composition of the MIM layers is described in the main manuscript.

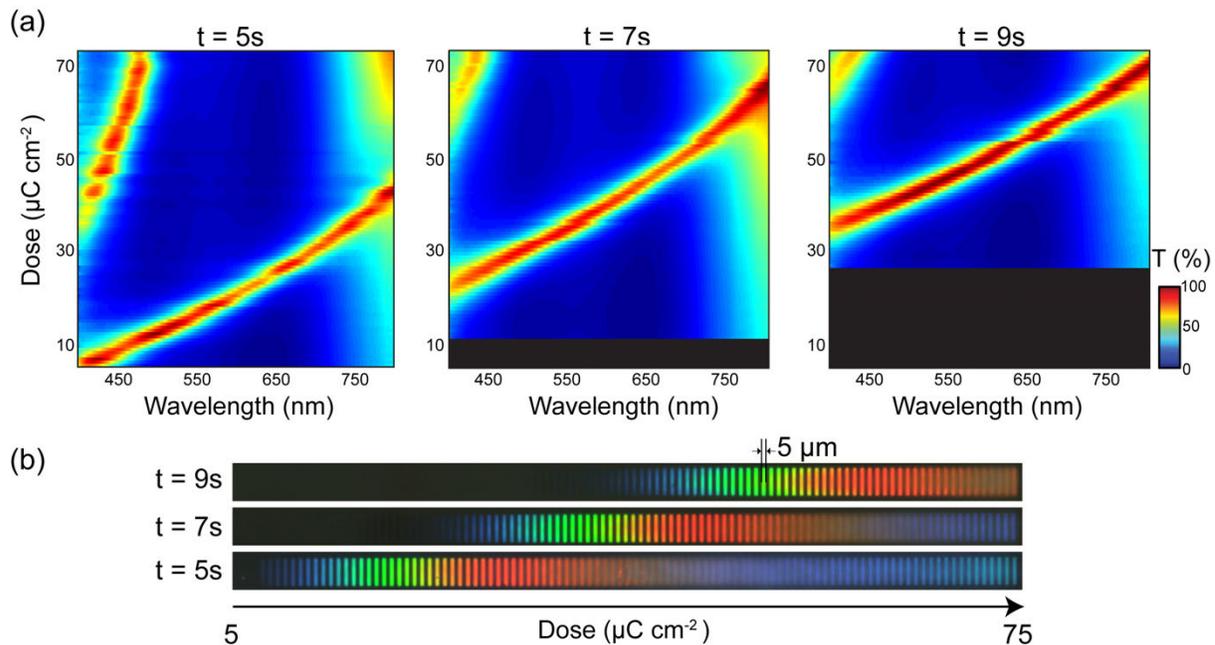

**Supplementary Figure 4. The effect of dose variation and development time.** (a) Measured transmission spectra, as a function of exposure dose (µC cm⁻²), for three different development times. (b) optical micrographs of a set of 5 µm wide rectangles with varying exposure dose for three development times. MaN-2405 resist is used along with full concentration AZ 726 MIF developer; as described in the *Methods*.

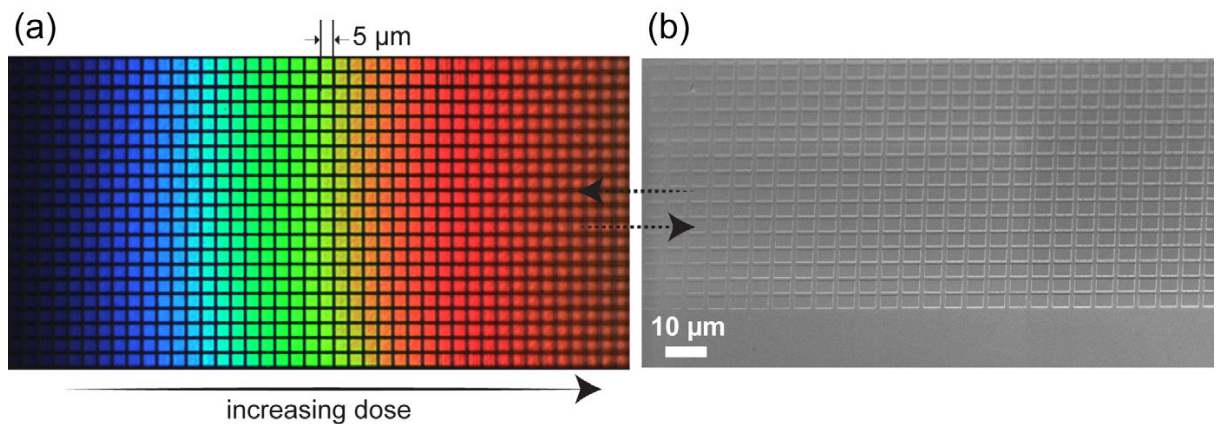

**Supplementary Figure 5. Increasing exposure dose with 5 µm pixels.** (a) Optical micrograph (transmission) of an array of 5 x 5 µm squares (pixels) which linearly increase in exposure dose (left-to-right). The increase in dose results in a thicker remaining resist (insulator in MIM cavity) hence red-shifted transmission characteristics. (b) An SEM micrograph of the same area in (a) in which the variation in height (from left-to-right) can be observed.

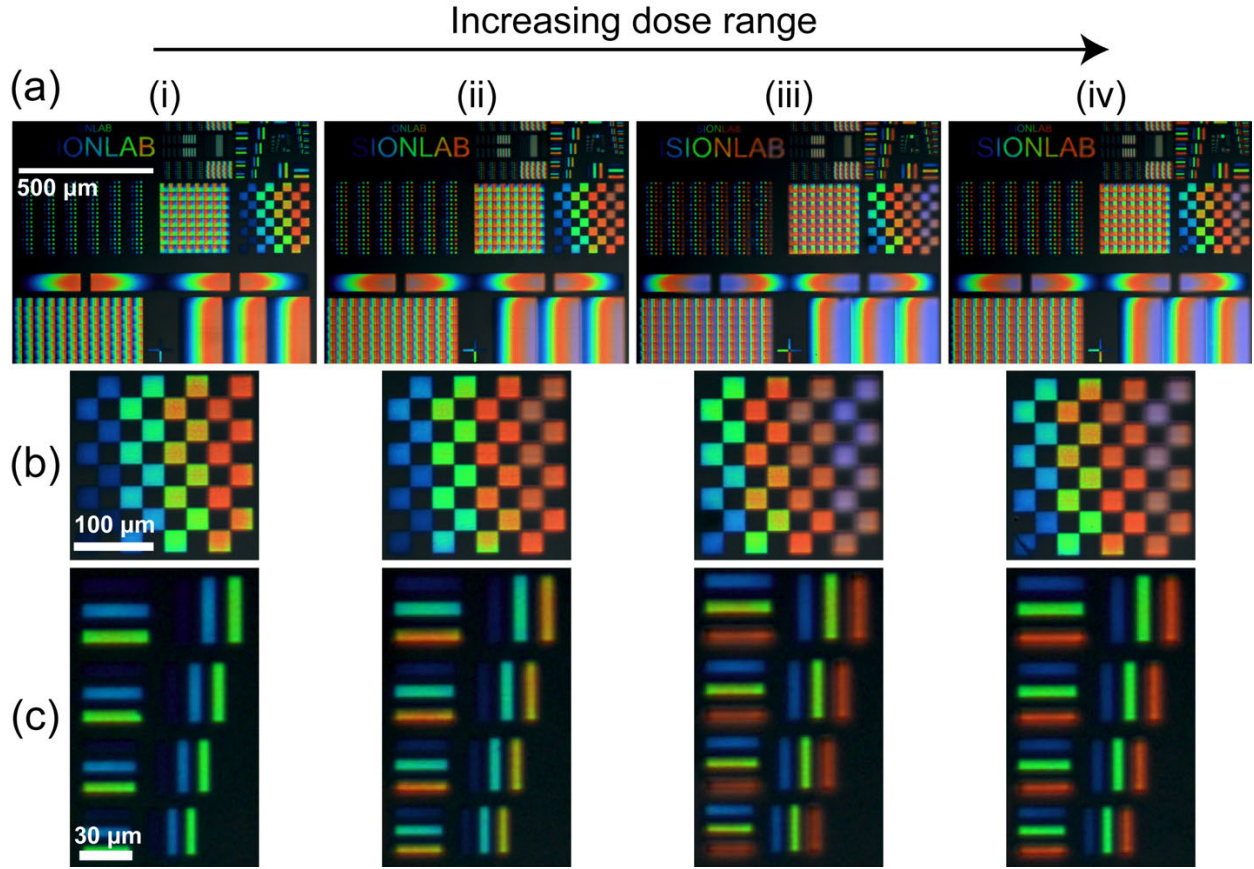

**Supplementary Figure 6. Dose/Resolution test with increasing dose limits.** (a) A custom resolution pattern in transmission under the optical microscope, composed of 100 different layers in which each layer has a distinct exposure dose between the (linear) range (i) 0.1–0.4 Cm$^{-2}$, with (ii – iv) increasing by +0.125 Cm$^{-2}$ each time. The sample is processed as described in the Methods. (b) and (c) show digitally zoomed in regions of specific areas of the dose/resolution test in (a). In (c), there are three different doses (one for each lined element), starting in (i): 0.16, 0.25, 0.34 Cm$^{-2}$. The figure effectively shows the change in spectral position of the FP-like mode of the MIM filter with varying exposure dose.

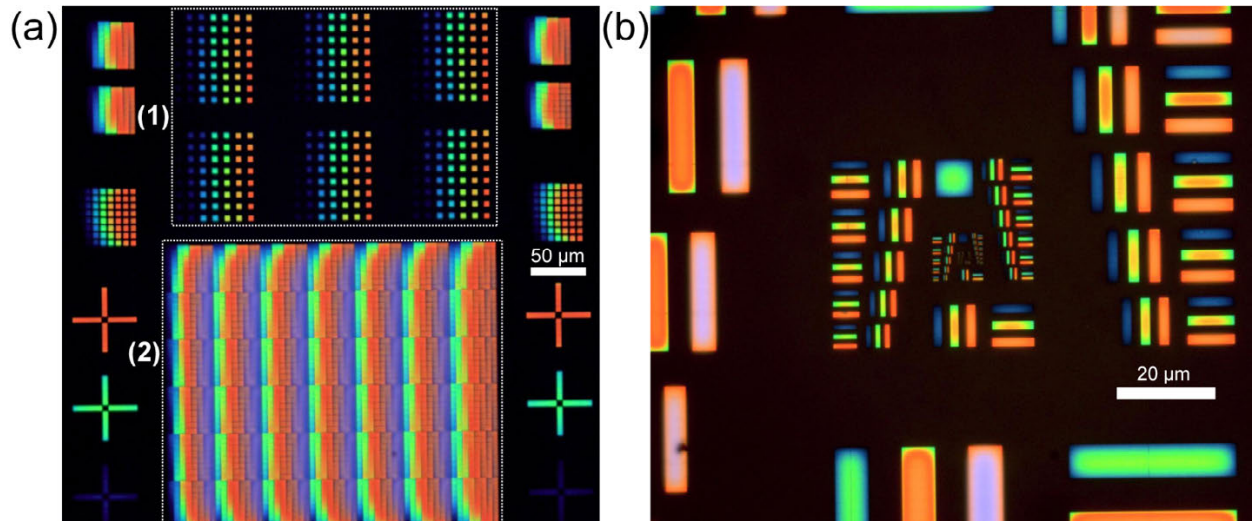

**Supplementary Figure 7. Visual example of the proximity effect using resolution test grids.**
(a) Dose test array in which the square pixels – shown in (1) and (2) – are the same size but vary in spacing. In (2) the proximity effect causes the final pixel thickness to be greater than in (1) due to the additional exposure from adjacent pixels. In (b), the proximity effect is highlighted in a USAF 1951 resolution target in which each element (in the line triplet) is given a different dose (hence different final wavelength). However, due to the increased 'dose density' in the center of the rectangle compared to the corners/edges, the final thickness and hence wavelength is different.

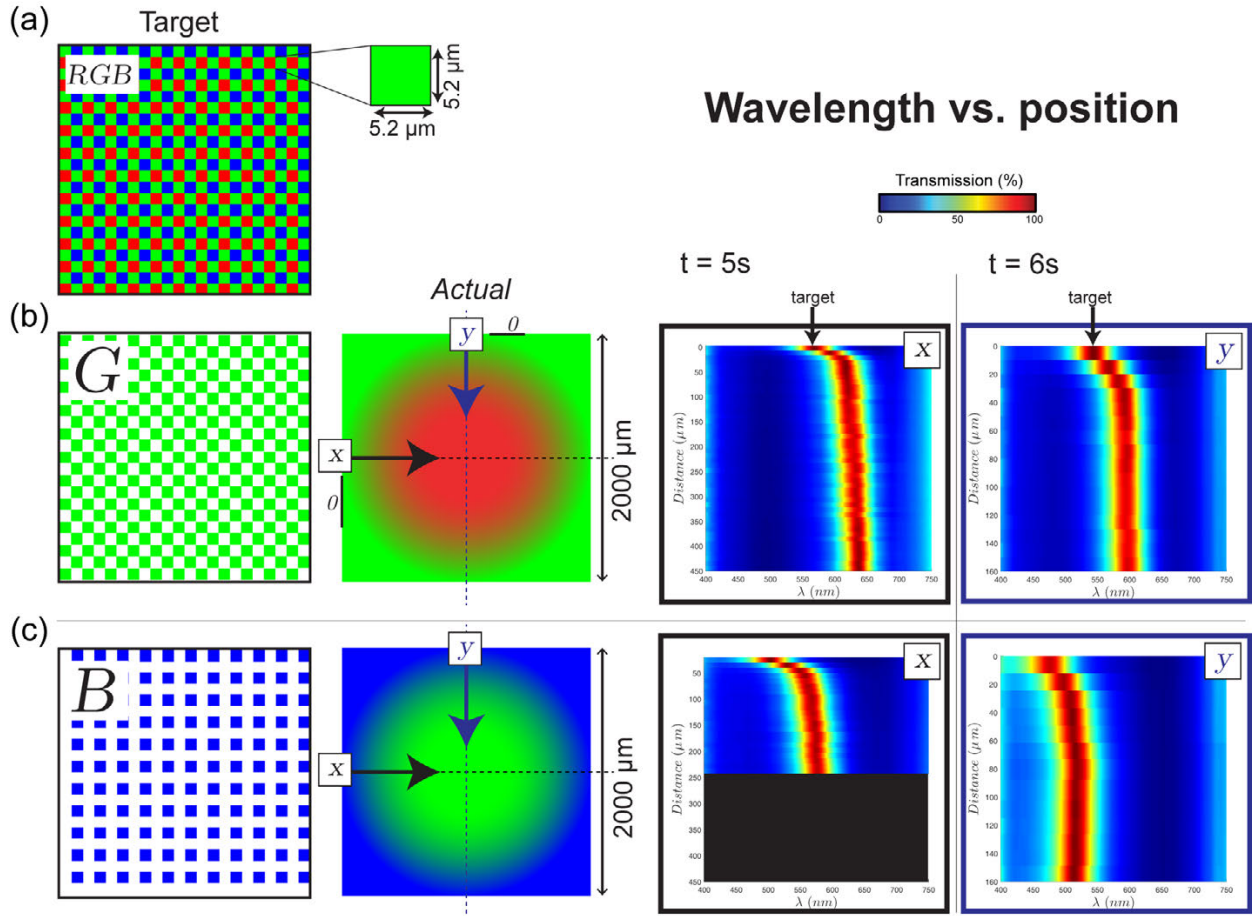

**Supplementary Figure 8. Proximity effect demonstration on RGB array.** (a) Desired final pattern: RGB CFA array. (b) and (c) are the desired green and blue channel mosaics respectively, with an exaggerated visual example of how the final color varies as a function of spatial position. Experimentally measured optical transmission of the respective wavelength channels in an RGB array as a function of distance from the edge of the patterned array, for two different development times.

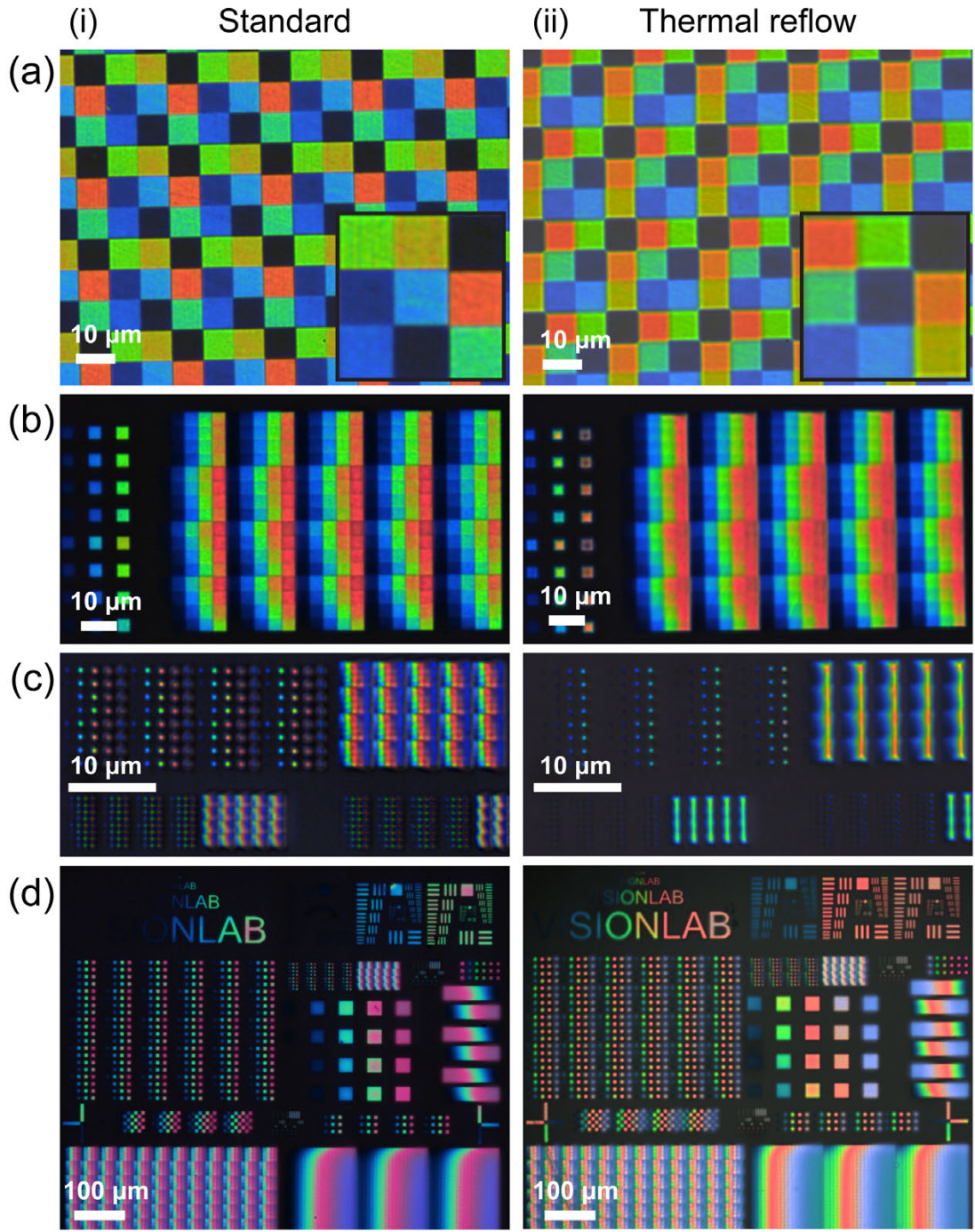

**Supplementary Figure 9. Resist reflow visualized.** Conditions: 100°C for 30s; with samples (a) 8.8 μm pitch pixel MSFA, (b) 3.75 μm pixel pitch pixels and (c) 900 nm (upper rows) and 460 nm (bottom rows) pixels. (d) Conditions: Pre-resist development thermal reflow, using 100°C for 1min; with custom resolution dose test. This is an example of a treatment to lower the dose applied to sample, as the temperature treatment here makes the resist harder to develop, and thus thicker.

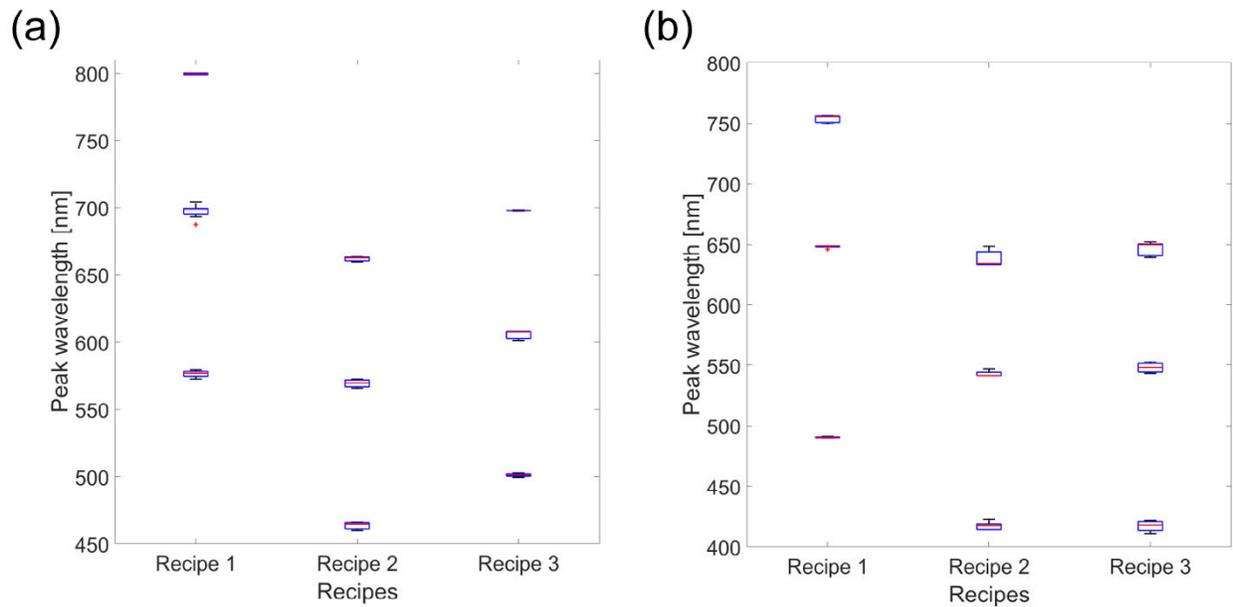

**Supplementary Figure 10. Peak wavelength box plots of RGB MSFAs for three different recipes.** (a) and (b) correspond to two different dose sets for red, green and blue channels: i.e. $D_{B,G,R}$ = 0.2, 0.3, 0.4 Cm$^{-2}$ for (a), and $D_{B,G,R}$ = 0.15, 0.275, 0.375 Cm$^{-2}$ for (b). Within each sub-figure, there are three different recipes; described previously. For each recipe, ~16 filter pixels (spectra) per channel (e.g. RGB = 3) were recorded. The red central lines correspond to the median values, and the bottom and top edges of the box indicate the 25th and 75th percentiles respectively.

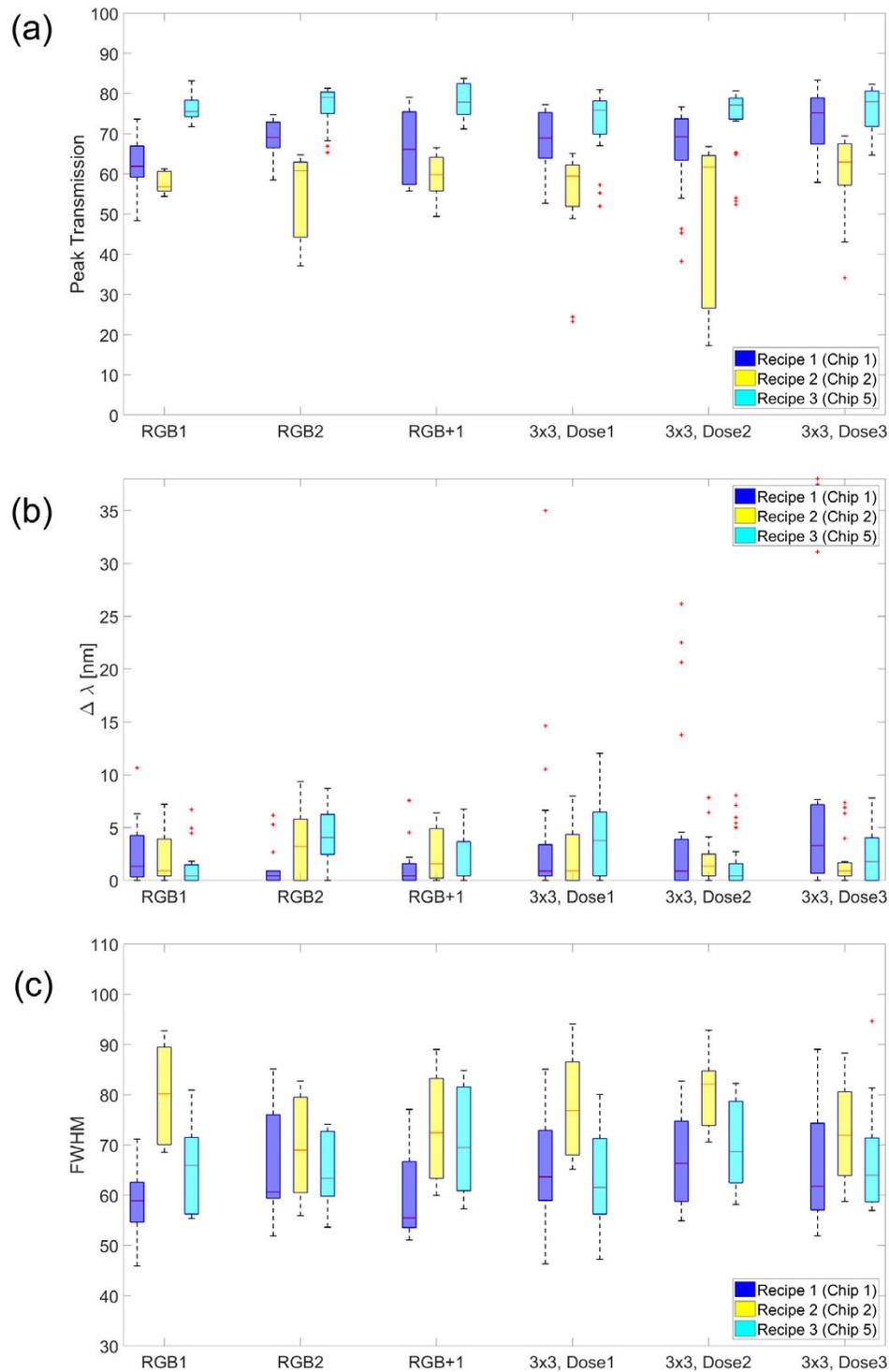

**Supplementary Figure 11. Box plots of the optical characteristics from a series of MSFA patterns from three different recipes.** (a) Peak transmission, (b) Peak wavelength shift, $\Delta\lambda$, from the average (i.e. $\Delta\lambda = |\lambda - \lambda_{av}|$) and (c) FWHM. The red central lines correspond to the median values, and the bottom and top edges of the box indicate the 25th and 75th percentiles respectively. For every CFA, several unit cells in the middle of each array were picked randomly and the spectrum of each pixel was recorded. For the fewer band (<4) MSFAS, ~12 spectra were recorded for each recipe. For the larger band MSFAs, 18–27 spectra were recorded for each recipe.

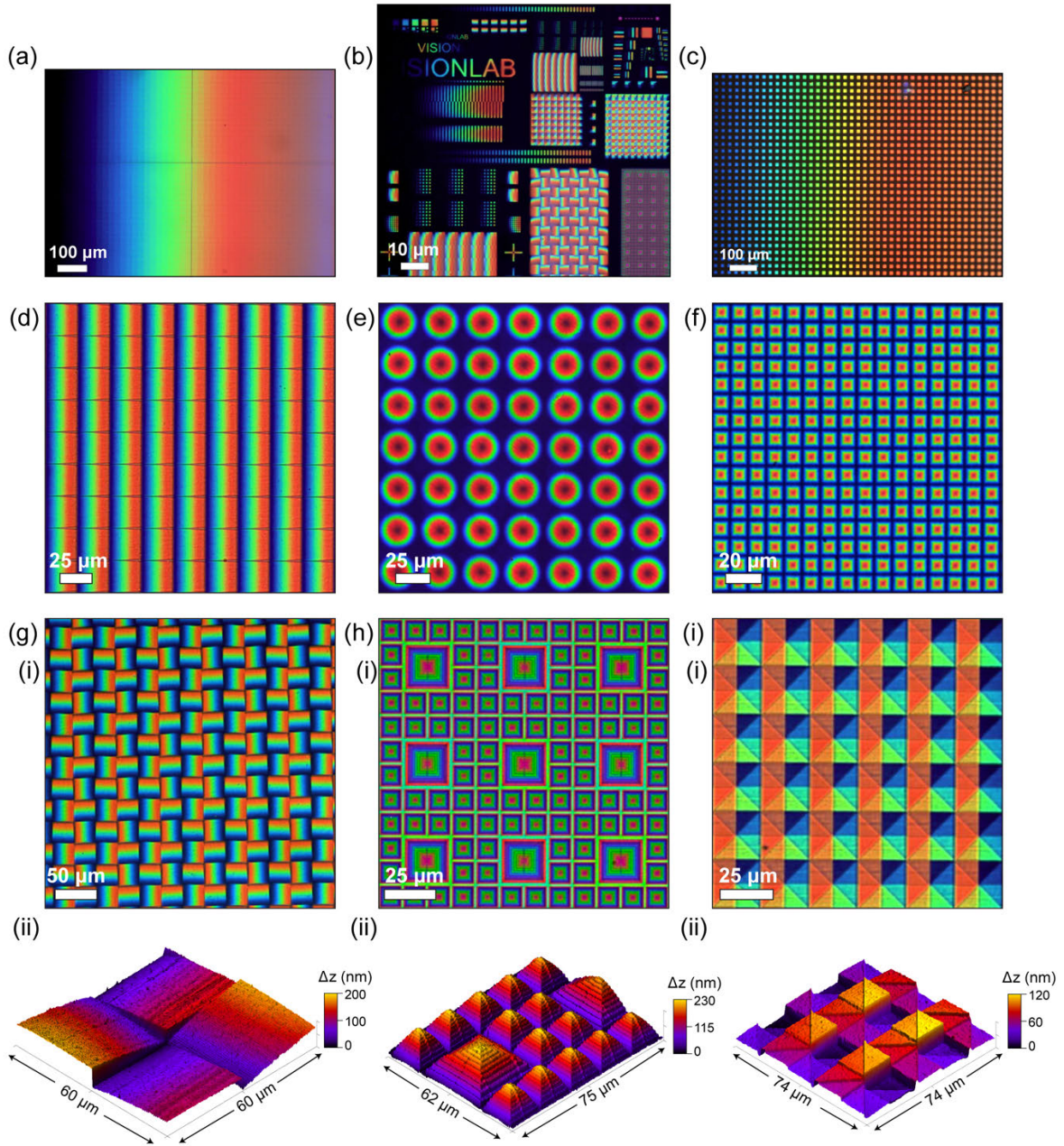

**Supplementary Figure 12. A variety of designs illustrating the versatility of the approach in this work.** This figure illustrates the customizability and power of the framework presented in this work, as all of the different designs chip (a—i) are fabricated with a single lithographic step on the same chip with the same materials (using the recipe in the Methods; main manuscript).

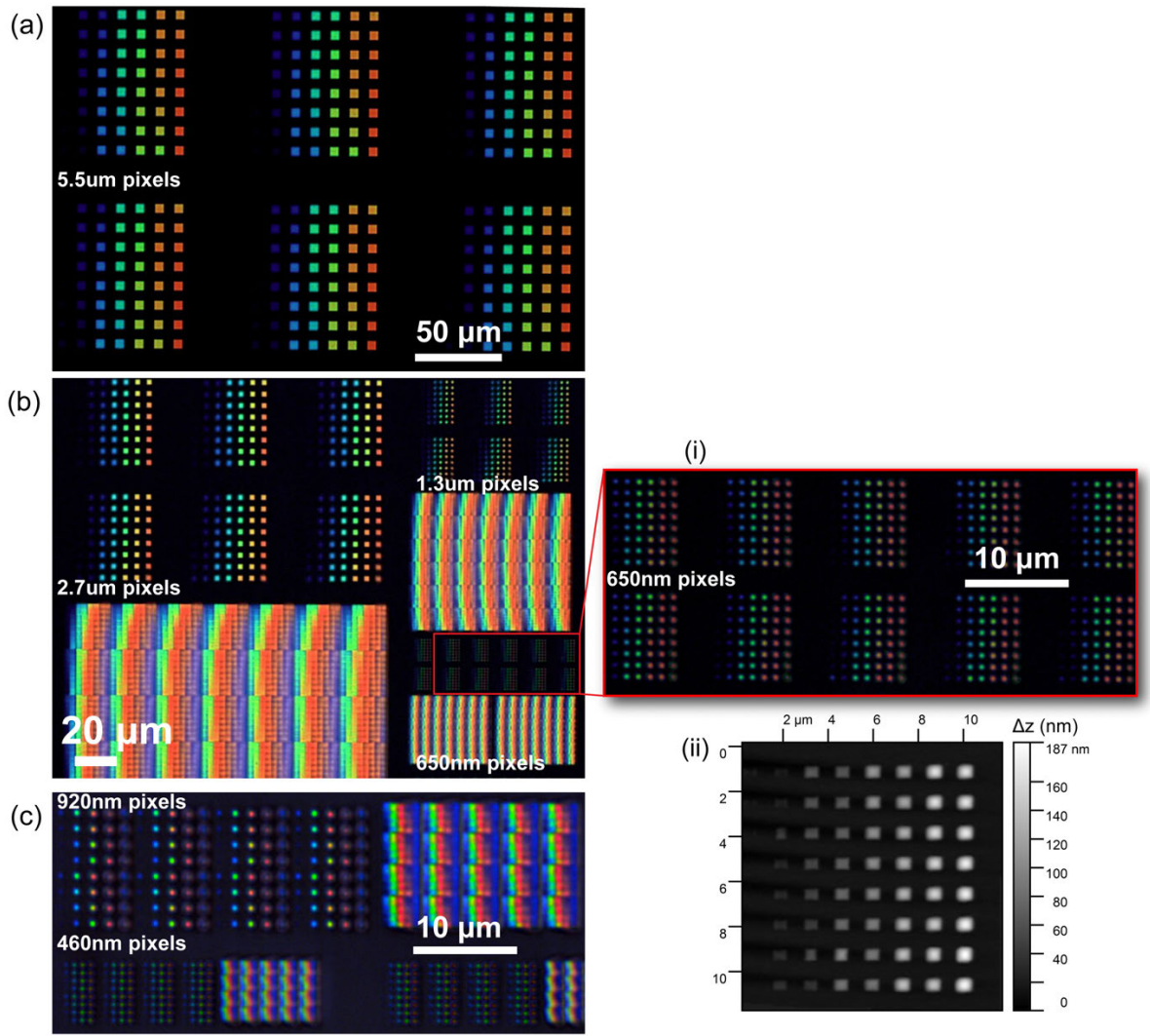

**Supplementary Figure 13. Resolution of pixels in this study.** A range of square pixels based on the grayscale MIM approach in this work under the optical microscope in transmission. (a) 5.5 µm lateral size pixels; (b) a range of pixel sizes from 2.7 µm, 1.3 µm and 650 nm; (i) is obtained with a higher magnification objective (100x) and (ii) is an AFM micrograph of one of the 650 nm pixel size sub-arrays in (i). (c) 920 nm and 460 nm pixel sizes. These resolution dose tests are fabricated as described in the Methods.

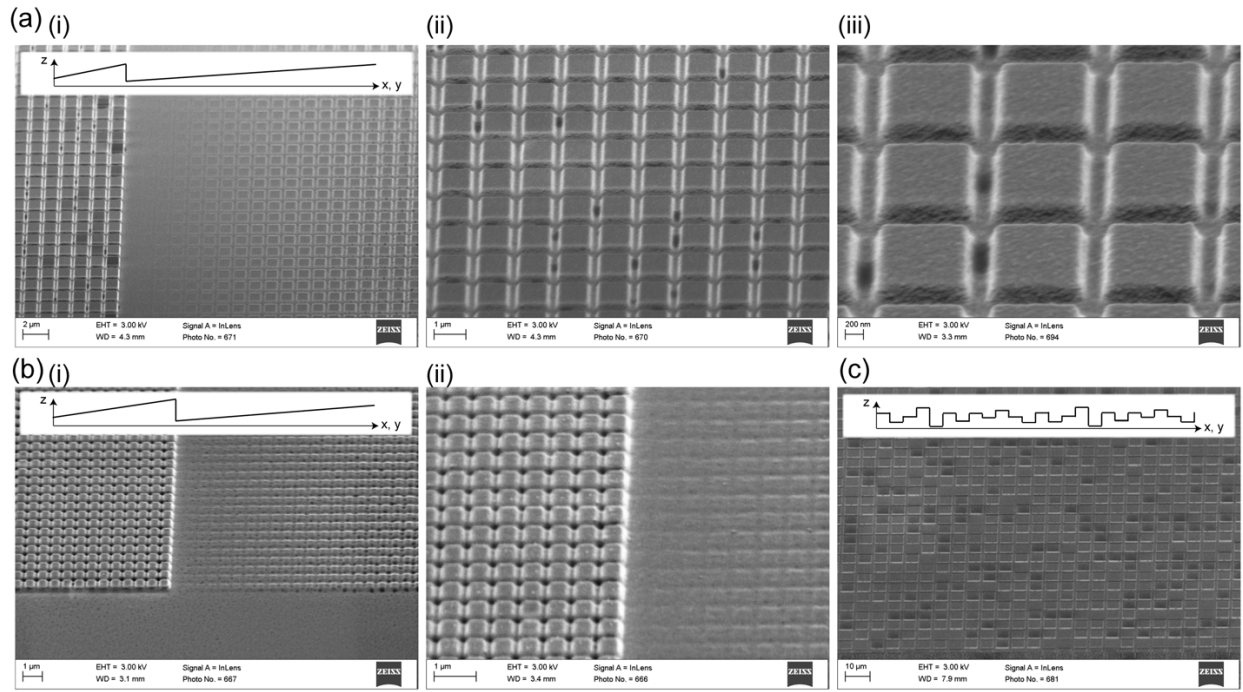

**Supplementary Figure 14. SEM micrographs of MIM pixel arrays.** (a) 1μm pixel array in which the dose (and thus final insulator thickness) varies in 1D. This 1D dose variation repeats (highlighted by the *inset*), with the transition shown in (i) and increasing magnifications shown in (ii) and (iii). (b) As in (a), but with a 500 nm pixel array. (c) A random dose array for 1 μm pixels.

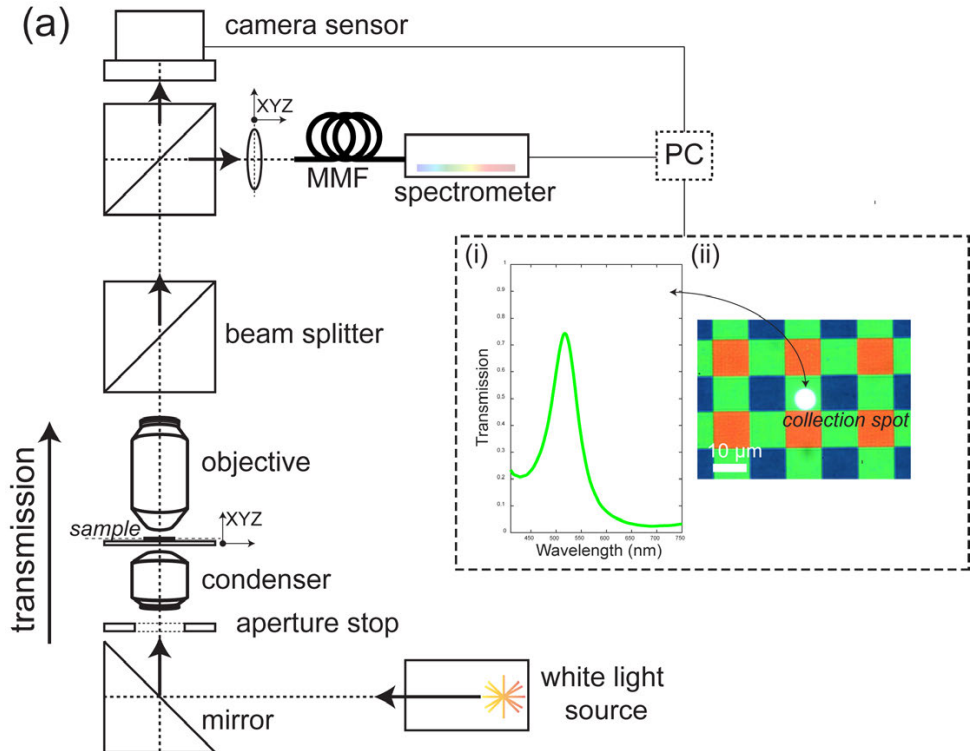

**Supplementary Figure 15. Optical characterization schematic.** The MSFAs are optically characterized using a modified Olympus BX-51 microscope: System schematic in (a). A 100W halogen lightbulb (V100WHAL-L; Philips 5761) illuminates the samples through a condenser (total light controlled via an aperture stop). A range of Olympus objectives (5—100x) are used for imaging the sample surfaces. Transmission is normalized through a glass cover slip (which the samples rest upon) and a 525 µm thick borosilicate glass (bright state). The two output paths are to a camera sensor and spectrometer to obtain magnified images of the sample and its corresponding transmission spectra respectively. (i) and (ii) show the information obtained from this arrangement: the transmission spectrum (i) and image of the sample surface in transmission (ii). The spectra are obtained using an Ocean Optics HR2000+ spectrometer and Ocean Optics Oceanview software

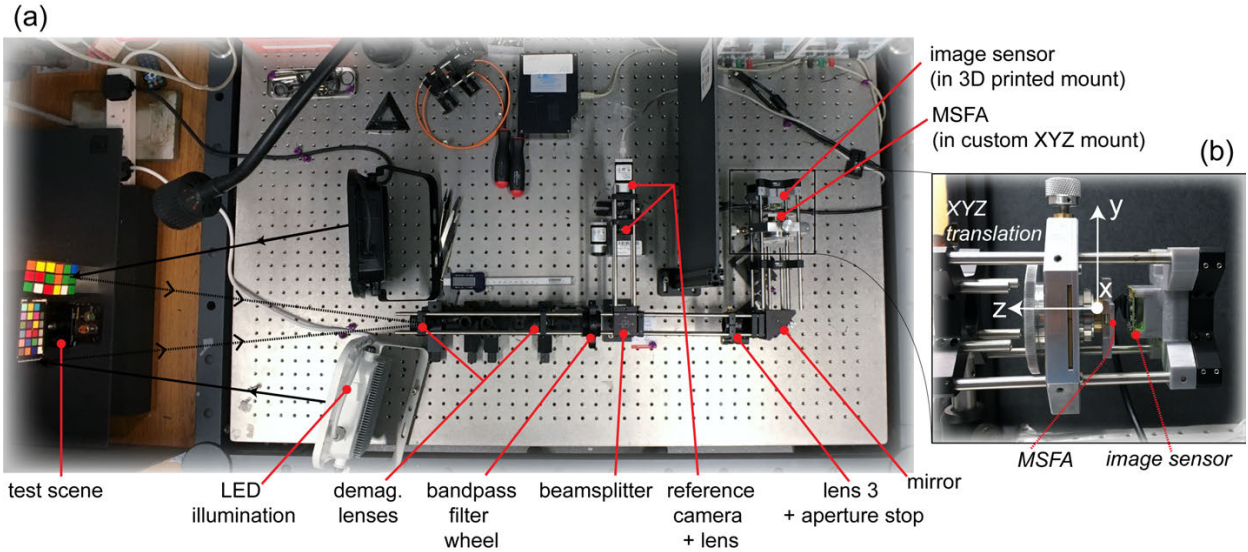

**Supplementary Figure 16**. **Optical imaging setup**. (a) Photograph of the optical system used for imaging experiments, in which a multispectral test scene (color chart and Rubik's cube here; left) are illuminated with 2x white light LED sources, the reflected light is de-magnified then imaged onto a monochrome image sensor through a custom MSFA. (b) shows a different viewing angle on the MSFA located in an in-house built XYZ-translation cage

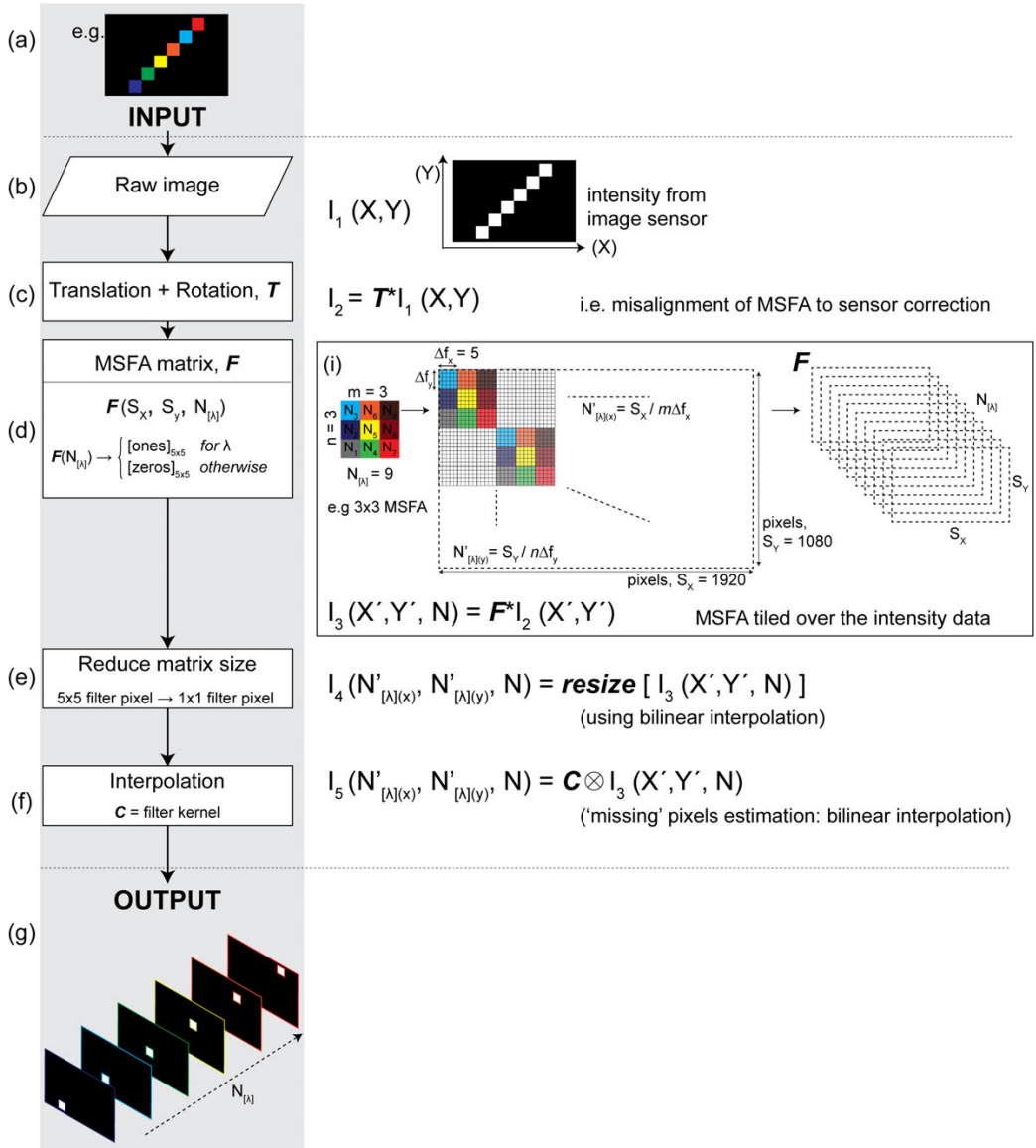

**Supplementary Figure 17. Algorithm flow chart used to demosaic the images acquired through the MSFA.** (a) Multispectral test scene is imaged onto the image sensor and a 2D-intensity matrix (0—255) is recorded; the raw image (b). A homogenous transformation matrix (T), incorporating a rotation by an arbitrary angle followed by a linear translation, is applied to the raw image (c), to account for the misalignment of the MSFA to the image sensor. This transformed matrix is then multiplied by a MSFA matrix (with N-bands), (d), which decomposes the 2D-intensity matrix into N x 2D matrices (one for each wavelength band). The mapping of filter pixel to image sensor is then taken into account in the MSFA matrix. This matrix is then reduced in size by a factor of 5 in each dimension (through bilinear interpolation). Finally, each channel is interpolated (d) by a pixel specific filter kernel (akin to Bayer filter demosaicing) to estimate the missing pixels between actual data. The output (g) is then N x 2D matrices: one for each wavelength band.

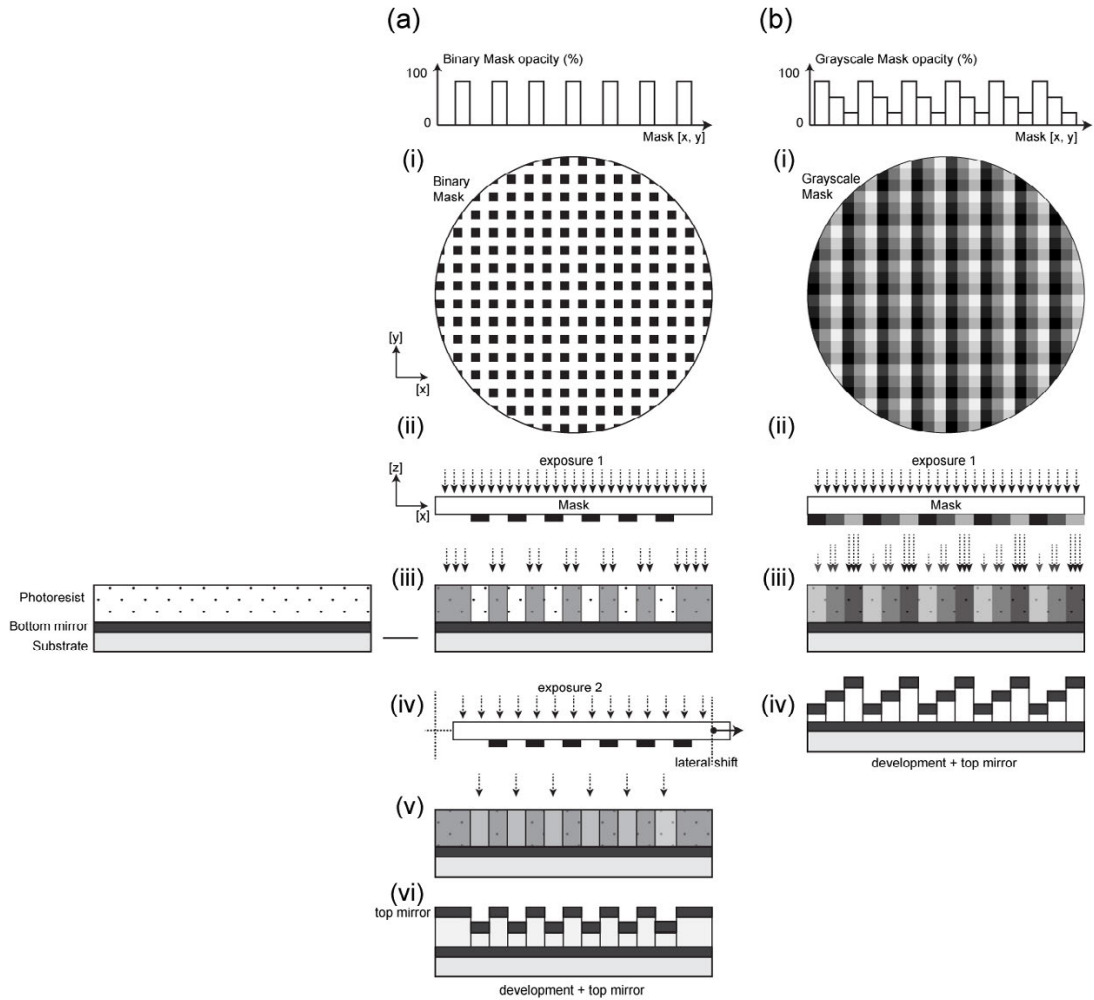

**Supplementary Figure 18. Concept 1: Translation of grayscale-to-color MSFA fabrication to photolithography via photomask approach.**

**(a) *Binary mask approach*.** (i) Opacity function of an example binary PL mask as a function of lateral coordinates [x,y]. (ii) Cross-section view [x,z] of the same binary mask with a flood exposure. (iii) Exposure 1: The areas in (ii) which are opaque 'block' the light, whereas the transparent areas pass the light–as with conventional mask-based PL. (iv) The same mask is laterally shifted (or the ability to use an additional binary mask is possible here). (v) Exposure 2 is performed, where exposure 2 ≠ exposure 1. *Note*: this step can be repeated for arbitrary designs, whereby each exposure yields a different final resist thickness. Thus, on the schematic there are two alternating parts of the resist which correspond to different exposure doses. (vi) Development: The resist is developed producing alternating thicknesses. Note: these thicknesses and pattern vary depending on the mask/s used. The top mirror is deposited creating a cavity (MIM geometry) and spatially variant filters are subsequently produced.

**(b) *Grayscale mask approach.*** (i) Opacity function of an example grayscale PL mask as a function of lateral coordinates [x,y]. The different graylevels correspond to different levels of attenuation of the light. (ii) Cross-section view [x,z] of the same grayscale mask with flood exposure. (iii) Exposure 1: The areas in (ii) determine the attenuation level of the light, thus the imparted dose profile and finally the resulting resist thickness. i.e. the more transmissive the region, the more light through, the thicker the final resist thickness and more red shifted the spectral response. (iv) Development: The resist is developed producing a grayscale thickness profile. The top mirror is deposited which creates a cavity (metal-insulator-metal geometry or otherwise) and spatially variant filters are subsequently produced.

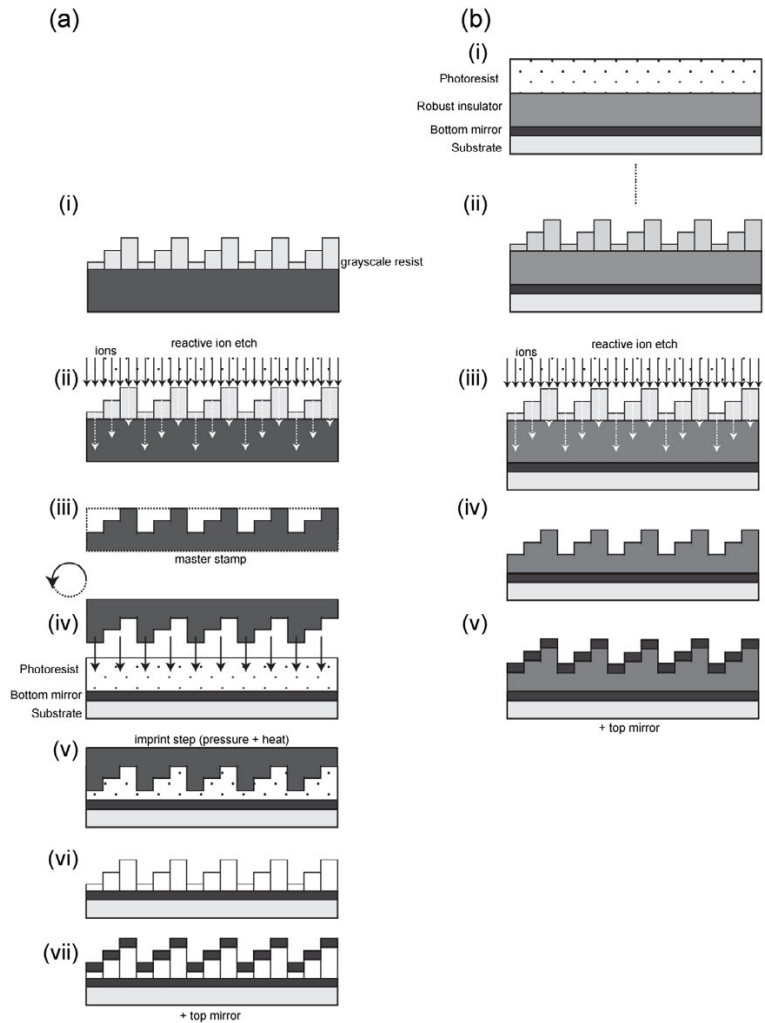

**Supplementary Figure 19. Concept 2: Translation to photolithography via etching.**

**(a) *Grayscale master stamp.*** (i) Grayscale resist pattern atop of the master stamp material. This has been produced through any means (EBL / PL), it is not important which is used here. (ii) Etching step (RIE or otherwise): The resist and master stamp material is etched anisotropically, whereby etch thickness into the stamp material is determined through the resist thickness. If RIE is used, this step is performed using heavy ions bombarding the sample. RIE and imparting the pattern is a known process. (iii) The resultant master stamp is produced. (iv) Rotating the stamp and bringing into contact with a polymer (photoresist or otherwise) atop of a mirrored substrate. (v) Imprinting / moulding step: Stamping into the polymer, and incorporating a pressure + heat step (not exclusively necessary). (vi) Removing the master stamp leaving the imparted master stamp pattern into the polymer. (vii) The top mirror is deposited which creates a cavity (metal-insulator-metal geometry or otherwise) and spatially variant filters are subsequently produced.

**(b) *Robust insulator.*** (i) i.    The starting material includes an insulator layer atop the bottom mirror, which is not the photoresist to be patterned. It has been deposited prior to the lithographic step. An example of a good material here is SiO2 (e.g. quartz). (ii) The lithographic step (EBL / PL) has been performed to produce a grayscale pattern into the photoresist. (iii**)** Etching step (RIE or otherwise): The resist and 'robust' insulator material is etched anisotropically, whereby etch thickness into the insulator is determined through the resist thickness. If RIE is used, this step is performed using heavy ions bombarding the sample. RIE and imparting the pattern is a known process.  (iv) The final 'robust' insulator atop the bottom mirror layer and substrate, which has taken the form of the grayscale resist thickness profile. (v) The top mirror is deposited which creates a cavity (metal-insulator-metal geometry or otherwise).